\documentclass[acmsmall]{acmart}

\usepackage{xspace}
\usepackage[capitalise]{cleveref}
\usepackage[colorinlistoftodos,prependcaption,textsize=tiny, loadshadowlibrary]{todonotes}
\usepackage{subcaption}
\usepackage{listings}
\usepackage{tabularx}
\usepackage{booktabs}
\usepackage{multirow}
\usepackage{changebar}
\usepackage[normalem]{ulem}
\usepackage{xcolor}

\lstdefinelanguage{TypeScript}{
	keywords={const, let, var, function, return, if, else, for, while, new, true, false, test, expect},
	sensitive=true,
	keywordstyle=\color{blue}\bfseries,
	ndkeywords={type, interface, implements, public, private, string, boolean, number, Date},
	ndkeywordstyle=\color{teal}\bfseries,
	identifierstyle=\color{black},
	commentstyle=\color{gray}\ttfamily,
	stringstyle=\color{orange}\ttfamily,
	morecomment=[l]{//},
	morecomment=[s]{/*}{*/},
	morestring=[b]",
}

\crefname{lstlisting}{listing}{listings}
\Crefname{lstlisting}{Listing}{Listings}

\graphicspath{{./}{./img/}}

\newcommand{\GF}[1]{\textcolor{green}{\textsf{\textbf{GF}:~#1}}}

\newcommand{\toolname}{\mbox{IntelliGame}\xspace}
\newcommand{\universityname}{\mbox{Politecnico di Torino}\xspace}
\newcommand{\control}{\emph{control}\xspace}
\newcommand{\treatment}{\emph{treatment}\xspace}

\newcommand{\summary}[2]{%
	\begin{center}%
		\colorbox{gray!20}{%
			\parbox{.97\linewidth}{%
				\textbf{\textsf{Summary (\textit{#1})}:}~%
				#2%
			}%
		}%
	\end{center}%
}

\newtoggle{showrevisions}

\togglefalse{showrevisions}

\iftoggle{showrevisions}{%
	\definecolor{reddish}{rgb}{0.85, 0.30, 0.0}
	\newcommand{\RA}[1]{\textcolor{reddish}{\textsf{#1}}}
	\newcommand{\RAremoved}[1]{\textcolor{reddish}{\sout{#1}}}
	
	\definecolor{bluish}{rgb}{0.0, 0.0, 0.90}
	\newcommand{\RB}[1]{\textcolor{bluish}{#1}}
	\newcommand{\RBremoved}[1]{\textcolor{bluish}{\sout{#1}}}
	
	\definecolor{purplish}{rgb}{0.8, 0.0, 0.8}
	\newcommand{\RC}[1]{\textcolor{purplish}{#1}}
	\newcommand{\RCremoved}[1]{\textcolor{purplish}{\sout{#1}}}
}{%
	\newcommand{\RA}[1]{#1}
	\newcommand{\RB}[1]{#1}
	\newcommand{\RC}[1]{#1}
	\newcommand{\RAremoved}[1]{}  
	\newcommand{\RBremoved}[1]{}  
	\newcommand{\RCremoved}[1]{}  
}

\newcommand{\comparison}[1]{
	\vspace{2mm}
	\noindent
	\fbox{%
		\parbox{.97\linewidth}{%
			#1
		}%
	}%
	\vspace{2mm}
}

\AtBeginDocument{%
	}

\setcopyright{acmlicensed}
\acmDOI{10.1145/3728983}
\acmYear{2025}
\acmJournal{PACMSE}
\acmVolume{2}
\acmNumber{ISSTA}
\acmArticle{ISSTA106}
\acmMonth{7}
\received{2024-10-31}
\received[accepted]{2025-03-31}

\begin{document}
	
	\title{Gamifying Testing in IntelliJ: A Replicability Study}
	
	\author{Philipp Straubinger}
	\authornote{Both authors contributed equally to this research.}
	\orcid{0000-0002-9265-5789}
	\affiliation{%
		\institution{University of Passau}
		\city{Passau}
		\country{Germany}
	}
	\email{philipp.straubinger@uni-passau.de}
	
	\author{Tommaso Fulcini}
	\authornotemark[1]
	\orcid{0000-0001-8765-6501}
	\affiliation{%
		\institution{Politecnico di Torino}
		\city{Torino}
		\country{Italy}
	}
	\email{tommaso.fulcini@polito.it}
	
	\author{Giacomo Garaccione}
	\orcid{0000-0001-7254-9578}
	\affiliation{%
		\institution{Politecnico di Torino}
		\city{Torino}
		\country{Italy}
	}
	\email{giacomo.garaccione@polito.it}
	
	\author{Luca Ardito}
	\orcid{0000-0002-0501-7886}
	\affiliation{%
		\institution{Politecnico di Torino}
		\city{Torino}
		\country{Italy}
	}
	\email{luca.ardito@polito.it}
	
	\author{Gordon Fraser}
	\orcid{0000-0002-4364-6595}
	\affiliation{%
		\institution{University of Passau}
		\city{Passau}
		\country{Germany}
	}
	\email{gordon.fraser@uni-passau.de}


	\begin{abstract}
		
		Gamification is an emerging technique to enhance motivation and performance in traditionally unengaging tasks like software testing. Previous studies have indicated that gamified systems have the potential to improve software testing processes by providing testers with achievements and feedback. However, further evidence of these benefits across different environments, programming languages, and participant groups is required.
		This paper aims to replicate and validate the effects of \toolname, a gamification plugin for IntelliJ IDEA to engage developers in writing and executing tests. The objective is to generalize the benefits observed in earlier studies to new contexts, i.e., the TypeScript programming language and a larger participant pool.
		The replicability study consists of a controlled experiment with 174 participants, divided into two groups: one using \toolname and one with no gamification plugin. The study employed a two-group experimental design to compare testing behavior, coverage, mutation scores, and participant feedback between the groups. Data was collected through test metrics and participant surveys, and statistical analysis was performed to determine the statistical significance.
		Participants using \toolname showed higher engagement and productivity in testing practices than the control group, evidenced by the creation of more tests, increased frequency of executions, and enhanced utilization of testing tools. This ultimately led to better \RA{code} implementations, highlighting the effectiveness of gamification in improving functional outcomes and motivating users in their testing endeavors.
		The replication study confirms that gamification, through \toolname, positively impacts software testing behavior and developer engagement in coding tasks. These findings suggest that integrating game elements into the testing environment can be an effective strategy to improve software testing practices.
		
	\end{abstract}

	
	\maketitle
	
	\section{Introduction}
	In software development, ensuring the quality and reliability of software products is a key priority. Software testing plays a crucial role in this process, serving as a core element of the software development lifecycle~\cite{tuteja2012research}. It helps identify defects, improve functionality, and ensure that applications meet their intended requirements.
Despite its importance, however, software testing is often seen as one of the less appealing phases of development, which can lead to testers feeling undervalued or disengaged~\cite{10301252}. This lack of motivation can hinder productivity and negatively impact the overall quality of the software produced~\cite{Motivation}.

To address these challenges, recent literature has explored the concept of \textit{gamification}, a technique that applies game-design elements in non-game contexts~\cite{Deterding}. Across various aspects of Software Engineering gamification has shown promising results in terms of engagement and performance~\cite{10.1145/3582273}.
There have been several attempts to incorporate gamification into the software testing process, both to enhance learning (especially in academic settings) and to improve the practical execution of testing. Gamified tools and environments have been applied to a range of testing activities, including unit testing and GUI testing~\cite{10375892, 10190463, 9787968}, test creation, execution, and maintenance~\cite{7528958, original_experiment, straubinger2024gamifying}.

Most studies in this area have focused on exploring new tools and conducting preliminary evaluations to validate their effectiveness and user experience (UX) in small-scale contexts~\cite{10.1145/3582273}. So far, these gamified approaches have largely been developed independently, with limited efforts to generalize the results across different contexts. 
However, there is a need for consolidation of the findings, requiring an effort to replicate the existing validation studies in other contexts, to achieve higher confidence in the obtained results and move towards well-established understanding. 

With this goal in mind, we conducted a replication study to assess one of the latest gamified tools for unit testing: \toolname, a gamified plugin for the IntelliJ IDEA Integrated Development Environment (IDE) that rewards testers with achievements for good testing practices~\cite{original_experiment}. Given the promising results of the initial evaluation, this tool merits further exploration, as limited participation may restrict the generalizability of observed benefits.

This paper presents our attempt to adapt \toolname for broader use, validating it through a controlled empirical experiment aimed at generalizing its effectiveness.
The key contributions of this paper are as follows:
\begin{itemize}
    \item We present an empirical experiment to validate the existing \toolname tool using a two-group experimental design on a new programming language with a sample of 174 participants.
    \item We describe a statistical analysis of participants' behavior and performance in writing and executing tests.
    \item We present a comparison of the findings from the original study with those of our study.
\end{itemize}

Our replication confirms that gamification can effectively impact user behavior by encouraging the creation of more tests, more frequent test executions, and increased use of testing tools, like coverage reports and debugging tools. We observed that while both groups exhibited similar testing approaches, gamified participants wrote more tests, focused on finer details, and produced code with fewer failing tests of the reference test suite—implying a slightly better functional outcome. Additionally, we found that achievement levels in \toolname correlated positively with both tester motivation and the quality of test suites. 

These results align with several conclusions from the original study, specifically that gamification can increase testing frequency and improve user engagement with tools for quality assurance. However, our study adds nuances. For instance, we found that \toolname also impacts debugging behavior—a result not highlighted in the original study. Further introspection on test quality and individual achievements indicates that while gamification users created more detailed test cases, their increased focus on quantity occasionally led to higher failure rates and test smells, particularly in the area of date and time handling. This indicates that while gamification encourages thoroughness, the \toolname approach would benefit from additional or modified achievements aimed at quality to balance the focus on test quantity.
Further, unlike the original study, our participants in the \treatment group reported greater time pressure and showed a greater awareness of the quality of their tests compared to the \control group, likely due to the continuous feedback provided by \toolname. 

In summary, we successfully replicated the original study on a gamification plugin with achievement features in the integrated development environment, largely confirming its results.
	
	\section{Background}
	\label{sec:background}

\RAremoved{\textbf{Software Testing}: Testing is crucial in software development, with unit testing fundamental to good practices \cite{anand2019importance}. Despite its recognized importance, developers often overestimate their testing efforts, frequently claiming to invest more in testing than they do~\cite{Beller-testpatterns}. In reality, many write few or no tests, and automated testing remains uncommon in many projects~\cite{Beller-testpatterns, Beller-when}.}

\RAremoved{While costs and time constraints may contribute to this gap in testing, a significant factor is likely the lack of motivation among developers \cite{ke2010effects}.
A common strategy to encourage test writing and evaluate test quality is the use of code coverage metrics, which assess the portions of code executed by a test suite. Code coverage can include various metrics—such as statement, branch, line, and path coverage—that allow developers to see the proportion of code tested, often enhanced by visualizations and highlighting features~\cite{Zhu-unittest}.
However, considering code coverage alone as a quality metric is not entirely indicative of the effectiveness of the test suite, as it has been proven that there is no strong correlation between coverage and effectiveness in fault detection~\cite{coverage_correlated}.}

\RAremoved{An alternative metric to evaluate the test suite's quality is the mutation score, a concept inherited from the mutation testing technique.
The mutation testing technique is based on the injection of defects in the code purposely, verifying whether the existing test suite can detect these code mutations. The ratio between the amount of mutation detected and the total mutants injected is called mutation score~\cite{9402038}. }

\RAremoved{Although such metrics are a key indicator, they alone fail to motivate developers, necessitating the use of other motivating factors.
Prior research suggests that low motivation for testing arises because testing is often perceived as tedious, frustrating, and even stressful, deterring both students and professionals from investing time in it~\cite{10301252, Deak, Santos}.}

\subsection{Gamification in Software Testing}

Gamification is a technique adopted for increasing motivation and interest in unappealing tasks and is considered as the usage of game elements in non-recreational contexts~\cite{Deterding}. An additional benefit is that tasks performed in a gamified environment can produce better output compared to non-gamified equivalent tasks~\cite{Porto-mapping, Stol-gamification}.
Among the various game elements, the most commonly adopted ones in gamified approaches are points, leaderboards, badges, and awards~\cite{Barreto-review, deJesus-characterization}, with other possible elements such as avatars (a user's visual representation in the gamified system), progress bars (indicators of a user's progression toward completing a task), feedback (the game's reaction to the user's tasks and actions), achievements (specific rewards earned after completing tasks multiple times), and penalties (detrimental effects being applied after incorrect behavior, to deter further errors) being used depending on the application context.

Using gamification in software testing education has proven effective, boosting student motivation and engagement in what is often seen as a less appealing subject~\cite{deJesus-report, Yordanova}. Gamified methods have been successfully integrated into various aspects of software testing, with unit testing being a common area of focus~\cite{10.1145/3582273, deJesus-characterization}. Other areas that have seen gamification benefits include introductory courses~\cite{sheth2015gameful, Bell-halo}, testing tools~\cite{Clarke-tools, Fu-cybergamif}, exploratory~\cite{Lorincz-expreport, Costa-systematic, DBLP:conf/gamify/Straubinger024}, mutation~\cite{7528958}, and GUI testing~\cite{Cacciotto, Gerry}.

For instance, gamification was introduced into a software testing course~\cite{Blanco}, comparing student performance with those from a prior cohort using a traditional teaching approach. The gamified course led students to run more test suites and uncover more bugs than the non-gamified course. Similarly, GUI testing~\cite{10375892} was enhanced through coverage indicators, leaderboards, and scoring, comparing students in gamified and non-gamified groups. The gamified group showed greater page coverage and included more assertions in their test cases, highlighting gamification’s impact on engagement and thoroughness in testing.




\subsection{IntelliGame}

\begin{figure}[t]
    \centering
    \includegraphics[width=0.7\linewidth]{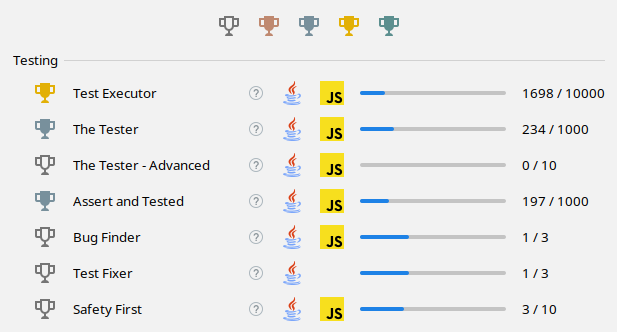}
    \vspace{-0.3cm}
    \caption{\RAremoved{IntelliJ} \RA{IntelliGame} window showing part of the achievements and their progress}
    \label{fig:achievements}
\end{figure}

\toolname is a plugin for the popular IntelliJ IDEA,\footnote{\url{https://www.jetbrains.com/idea/}} an Integrated Development Environment (IDE) that supports development in Java, Kotlin, and JavaScript. This plugin integrates gamification into the testing process through a straightforward yet effective game element: achievements~\cite{original_experiment}. \RC{\toolname tracks various testing-related activities within the IDE, such as running tests, using assertions, debugging, and improving test coverage. The system monitors user actions and awards good behavior with achievements based on predefined criteria. The implementation leverages IntelliJ’s event system using a publisher-subscriber pattern, ensuring accurate tracking of testing behaviors.}
\RC{The design ensures minimal workflow disruption by allowing developers to engage with gamification at their own pace. Progress is displayed through a user interface with trophies and progress bars (\cref{fig:achievements}), while notifications provide real-time feedback on achievements. }

\toolname offers a set of achievements across four key areas: testing, coverage, debugging, and test refactoring achievements:
\begin{itemize}
    \item \textbf{Testing achievements} reward developers for writing and running tests, encouraging frequent test execution, assertion triggers, and addressing failing tests.
    \item \textbf{Coverage achievements} motivate testers to use test coverage tools and improve coverage with each new test run.
    \item \textbf{Debugging achievements} recognize developers using the debugger of IntelliJ to fix bugs and resolve code issues.
    \item \textbf{Test Refactoring achievements} incentivize developers to enhance existing test code, such as by refactoring redundant code into helper functions.
\end{itemize}

Achievements are represented by different kinds of trophies indicating the tester's current level, a name, a brief description visible on hover, and a progress bar showing progress toward the next level.
Achievements provided by \toolname mimic important testing milestones, such as reaching a certain line, methods, and branch coverage running tests, or reward positive testing behavior such as measuring coverage or fixing tests. As soon as testers meet the requirements for a certain achievement, they are notified via a notification system inside the IDE.
\Cref{fig:achievements} illustrates the plugin's achievements window with all available achievements displayed.

\RC{Each achievement consists of multiple levels, with progress earned through actions such as running tests, using assertions, debugging, and improving test coverage. These achievement levels create a structured progression system that rewards developers for engaging in various testing activities. The boundaries for each level were set based on an analysis of typical developer workflows in a pilot study before the original study~\cite{original_experiment}, ensuring a balance between accessibility for beginners and meaningful challenges for experienced users. Levels unlock once these predefined thresholds are met, encouraging continuous learning and skill refinement in a motivating and structured way.}

\RA{The intended use case scenario for \toolname is mainly to support developers during their implementation and testing tasks, for example when onboarding new developers, or for improving quality-related habits for more experienced developers. Developers, while developing their codebase and writing and running unit tests should engage with the achievements, rewarding when these tests pass, when they are run with coverage, and when they are maintained by fixing code for failing tests. This scenario is particularly suitable for Test-driven development, where the coding is based on iterative refinement from a given test suite. It is also conceivable to integrate the plugin into academic courses to provide students with feedback while learning testing concepts. }

\toolname has been previously validated in a controlled experiment with 49 participants~\cite{original_experiment}. This first validation had the objective of assessing whether the tool can influence testing behavior and the quality of the resulting test suites and codebases. The participant's performance was measured in terms of test quality (number of tests written, code coverage, mutation scores), code functionality (number of tests passing against a reference test suite), and user experience.
To study the effects of these metrics, the participants were divided into four groups, each with a different treatment, and asked for the same implementation task: developing two Java functions and verifying the correctness of their implementation. 
The different treatments consisted of (1) a version of the gamification plugin providing only notifications corresponding to milestones, but no achievements to the users, (2) the \toolname plugin itself as previously described considering it as the treatment group, (3) a group using the gamification plugin but explicitly indicating to try to maximize the levels of the achievements and (4) a control group with a plugin with no effects, just collecting data.
Further information \RC{and more detailed explanations} regarding \toolname and this study are contained in the original study~\cite{original_experiment}.

\subsection{Replicability, Reproducibility, and Repeatability}

Empirical science is a cornerstone of the scientific method, as practical validation of theories is essential for scientific progress~\cite{Juristo2012}. Trust and reliability in empirical findings are critical to advancing theoretical and technical fields, with independent reproducibility being a fundamental requirement for accepting any empirical result as established~\cite{SHEPPERD2018120}.

Based on the ACM Replicability guidelines~\cite{ACM_guidelines}, there exists different concepts and granularities:
\begin{itemize}
    \item \textbf{Repeatability} is the ability of the original research team, under identical conditions, to confirm the experiment’s results across multiple trials.
    \item \textbf{Reproducibility} occurs when a different team, following the same measurement procedures, can achieve the same results and precision under identical operating conditions, regardless of location.
    \item \textbf{Replicability} is the ability of an independent team to obtain the same results and precision in a different location, using a different measurement system, across multiple trials.
\end{itemize}

The scientific community should ideally aim for the highest possible levels.
Our goal is to conduct a replicability study: As an independent team, we replicate the experiment with a similar, yet different experimental setup. The main differences are (1) a different and larger population of participants, (2) a different experimental object, with several non-trivial functions to implement, (3) a different target programming language, and (4) new research questions aiming to shed further light on the specific details and influences of the gamification aspects of \toolname.


	\section{A Replicability Study of the Initial \toolname Study}
	\label{sec:methodology}

In this section, we outline the experimental design according to guidelines for reporting software engineering experiments~\cite{1541818}. To ensure the validity of this description, we also used the checklist from the Empirical Standards for Software Engineering Research~\cite{EmpStandard}.

\subsection{Research Questions} 

The experiment aims to assess the impact of \toolname on its users. To achieve this, we established goals addressing different potential effects of \toolname on users, each linked to a specific research question (RQ).
Since the primary objective is to replicate an existing study, we adopted the original research questions from the initial validation paper~\cite{original_experiment}:
\begin{itemize}
\item \textbf{RQ1}: Does \toolname influence testing behaviour?
\item \textbf{RQ2}: \RA{How} does \toolname influence resulting test suites?
\item \textbf{RQ3}: Do achievement levels reflect differences in test suites and activities?
\item \textbf{RQ4}: Does \toolname influence the functionality of the resulting code?
\item \textbf{RQ5}: Does \toolname influence the developer experience?
\end{itemize}

The goals are as follows: first, to understand \toolname's influence on testing behavior and test quality; second, to evaluate whether achievement levels within \toolname affect the test suite; third, to assess the functionality of the assigned code base; fourth, to explore users' overall experience with \toolname. One aspect not considered in the original study is the quality of the tests and the importance of individual achievements themselves. Therefore, we add two new research questions to investigate test quality and achievements:

\begin{itemize}
\item \textbf{RQ6}: Do users of \toolname write high-quality tests?
\item \textbf{RQ7}: What are the most important achievements?
\end{itemize}

\subsection{Object Selection}
\label{sec:objsel}

The experimental object needed for the experiment had to be a real-world project to be used as a reference to assign students some programming tasks.
To select a suitable project for our needs, we began exploring a well-known database of JavaScript libraries in March 2023.\footnote{\url{https://web.archive.org/web/20230322130719/https://www.javascripting.com/}} We established specific inclusion criteria (IC) to determine if a project was eligible to be used as the experimental object.
The inclusion criteria were as follows:
\begin{itemize}
    \item The project must be an open-source artifact.
    \item The project must be implemented in JavaScript or TypeScript.
    \item The project should be a standalone library containing functions related to data types, easily understandable by students, rather than a development framework (e.g., Vue, or Angular).
    \item The project must include unit tests to verify the correctness of all its functions.
    \item The project documentation should be clear and comprehensive, detailing each function.
    \item The project must be well-maintained, with its last update occurring within three months.
\end{itemize}

We started browsing the database, focusing on the \textit{miscellaneous} category to find a project that met our criteria. We applied the inclusion criteria iteratively to each project on the list, which was ranked by popularity. After filtering, we identified Date Fns\footnote{\url{ https://github.com/date-fns/date-fns}} as a suitable choice.

The original Date Fns project features nearly 250 utility functions related to date data types. These functions range from simple tasks, such as comparing two dates, to more complex operations like transforming a date to conform to ISO or RFC standards.\footnote{ISO 8601 formats dates as year-month-day hour:minute:second.millisecond (e.g., 2022-09-27 18:00:00.000). The RFC standard includes the weekday, day, month, year, time in 24-hour format, and timezone code, such as Tue, 27 Sep 2022 18:00:00 EST.} This library is highly popular, used in millions of projects, and is well-maintained by more than 390 contributors, with over 1000 forks.

Since the original project utilized a different testing framework than the one the participants were familiar with, we needed to modify the configuration to suit our needs. Specifically, we manually converted the original test suite \(Ts\) from the Vitest\footnote{\url{https://vitest.dev/}} framework to the equivalent Jest format \(Ts'\). To ensure that our refactoring process did not introduce any errors, we executed the original and the refactored test suites, confirming that both produced no failures and the same mutation scores.

\subsection{Pilot Study}

Before conducting the main experiment, we carried out a pilot study to improve and refine our methodology. In 2023, we organized a preliminary attempt at \universityname to replicate the original experiment, confirm the feasibility, and fine-tune our approach.
In the pilot, we adapted \toolname to support JavaScript and TypeScript, using Jest as the testing framework. Out of the original 26 Java-based achievements, 19 were successfully implemented for JavaScript, covering testing, coverage, and debugging-related achievements.

\RA{Moving the focus of \toolname from Java to applications to JavaScript or TypeScript could have been done in two ways: by migrating the underlying IDE from IDEA to WebStorm, the web-specific IDE from JetBrains, or by adding  support for a different testing engine, namely Jest, in addition to JUnit. Since readjusting the plugin both to a different IDE and a different testing framework would require significantly more effort than simply adding the support for Jest, we decided to pursue the second path. In case a positive effect of gamification is measured also for testing in TypeScript, further research endeavors will be directed towards the complete porting of the plugin into WebStorm.}


The large-scale pilot study with 152 participants validated our methodology and highlighted areas for improvement, particularly regarding task complexity relative to time constraints. During the pilot, participants were initially tasked with implementing 23 JavaScript functions, but most could not complete half within the allotted time. Consequently, we scaled down the main experiment to 11 functions to better fit within time limitations.
Additionally, the object of study—Date-Fns, a JavaScript library—remained consistent across the pilot and main experiment. The 2023 course required students to complete a JavaScript group project, necessitating adjustments to convert the TypeScript-based Date-Fns project to JavaScript. However, in 2024, the course project shifted to TypeScript, easing compatibility and familiarity with Date-Fns for students.

Another critical insight from the pilot was the need for separate experimental sessions. In mixed sessions, students could identify their group assignment, which may have influenced performance. To counteract this, the main experiment hosted distinct sessions for each group, preventing participants from inferring differences in treatment.

For a complete description of the pilot study methodology and findings, see \cite{10.1145/3643796.3648459}.

\subsection{Experiment Design}

The primary objective of this study is to investigate how gamification influences users' testing behavior. To achieve this, we conduct a simple two-group experiment with different participants. The experiment involves using two versions of \toolname: one with gamification (the \treatment group) and one without (the \control group). The \treatment group receives the adapted version of \toolname, while the \control group is provided with a plugin that only collects data. \RA{Both plugin versions calculate the levels and store the same data. However, only the plugin in the treatment group displays achievements, shows progress, and sends notifications.}
Although the original experiment used four groups for the comparisons, the main analysis consists of comparing the treatment with the control group, which we focus on in our replication.

The controlled experiment involved a programming task in which participants were asked to develop code based on specific requirements and ensure its correctness. They are not given any restrictions on how to test their code.
We base the programming assignment on code extracted from an existing software project, similar to the approach taken in the original experiment, to enhance the ecological validity of the study. The task is designed to be challenging yet feasible within the allotted 150 minutes, both in terms of implementation and testing.

Participants are given a simplified version of the GitHub Date Fns project. We carefully selected 11 functions from the original project, prioritizing simplicity in terms of complexity and the number of code lines. This was intended to ensure that most participants could successfully develop and test these functions without feeling overwhelmed.\RA{The functions to be developed and tested include boolean functions checking date values such as \textit{isAfter}, \textit{isPast}, \textit{isWeekend}, getters for some values from the date value such as one that returns the number of days in a month of the given date, or the day of the month of a given date, namely \textit{getDaysInMonth} and \textit{getDate}. Another function required was to perform a sum of days to a given date: \textit{addDays}.}

The bodies of the functions are left empty, retaining only the requirement descriptions, function names, and parameters. The project includes complete documentation for the date class and functions, preventing participants from accessing online references or external suggestions. To verify the correctness of their implemented code, we provide two methods: a \texttt{main.ts} file that mirrors the setup of the original experiment and functions similarly to the Main class in Java, along with a Jest configuration containing blank test files for participants to complete and execute.

After the implementation phase, participants completed an online exit survey. The survey began with demographic questions about their studies, age, gender, and experience with TypeScript and Jest. The second page included general questions about the implementation and testing of the class. A third page, shown only to the \treatment group, gathered feedback on their experience with the plugin. Responses were based on a five-point Likert scale, using questions from the original study for consistency~\cite{original_experiment}. An optional free-text field allowed participants to add additional comments.

\subsection{Participant Selection}

We selected a sample of master’s degree students because they closely represent new hires undergoing onboarding~\cite{lorey2022storm} and because this group allowed us to recruit a large sample size effectively. 

To recruit participants, we issued a call for volunteers during a Software Engineering course at \universityname, where the experiment was conducted. We used an eligibility survey to gather applicants, which included basic demographic questions and information about their familiarity with the relevant work environment. Students were encouraged to participate in the study, with an additional course grade point offered upon successful completion. Success was based on participation and commitment rather than correctness or performance, with students free to withdraw from the experiment at any time.

We received 264 survey responses, with 232 students actually participating in the experiment. However, only 218 provided usable data, and of these, only 174 participants’ projects included the \texttt{TestReport} from \toolname. \toolname generates this report whenever the program is executed, debug mode is used, or tests are written or executed. Incomplete data was largely due to setup issues, failure to install the plugin, or lack of actual participation.

The participants were predominantly in their early twenties, with an average age of 23.4, and about one-quarter were female. All were students, though some had industry experience, albeit in the minority. Most had less than six months of experience with JavaScript and TypeScript, and nearly all had less than three months of experience with Jest. All students were actively enrolled in the Software Engineering course and were working on a Node.js project that involved testing, where they became familiar with the experimental framework, including the programming language, testing, and build environment. \RA{Testing experience for the students was limited to the Software Engineering course, however, at the time of the experiment the students had been actively engaged for a month in creating a test suite for their group project.}

\RB{Unfortunately, most of the students participating in the experiment had no working experience in the field, therefore we cannot fully generalize the findings for experienced testers. To explore the impact of \toolname on developers and testers more comprehensively, one potential approach would be to separate the data from students who possess prior industry experience or those who are simultaneously employed and studying. However, this analysis goes beyond the scope of this experiment, as the small number of individuals fitting these criteria without our participant pool would necessitate data collection across multiple years of experimentation.}

\subsection{Experimental Analysis}
\label{sec:expan}

The experiment's analysis compares metrics across two groups, the \control group without the use of \toolname, and the \treatment group using the adapted version of \toolname. To assess the significance of the differences between these groups, we use the exact Wilcoxon-Mann-Whitney test~\cite{10.1214/aoms/1177730491} to calculate \textit{p}-values, applying a confidence threshold of $\alpha=0.05$.

\subsubsection{RQ1: Does \toolname influence testing behaviour?}

We analyze testing behavior based on the number of (1) tests written, (2) tests run, (3) tests run with coverage enabled, and (4) instances of using debug mode to execute tests.

\subsubsection{RQ2: \RA{How} does \toolname influence resulting test suites?}

This research question focuses on assessing the quality of the test suites by measuring (1) the number of tests, (2) test coverage on both line and branch levels, and (3) the mutation score of the final test suites. The number of tests and test coverage are automatically collected and calculated during test execution using the Jest framework.\footnote{\url{https://jestjs.io/}} Mutation testing is conducted with StrykerJS (Stryker Mutator),\footnote{\url{https://stryker-mutator.io/}} a popular mutation testing engine in web development. 
To ensure accuracy, failing tests were excluded from the mutation analysis, as Stryker requires a fully passing test suite.

\subsubsection{RQ3: Do achievement levels reflect differences in test suites and activities?}

During the experiment, \toolname logged each user interaction, recording both the current levels achieved and the status of each achievement. We calculate the Pearson correlation~\cite{pearson1895vii} between these achievement levels and metrics such as line coverage, branch coverage, mutation score, and the number of tests to compare the two groups.

\subsubsection{RQ4: Does \toolname influence the functionality of the resulting code?}

To evaluate code functionality, we use the original project test suite—comprising 60 test cases that cover 95.83\% of the original implementation's lines—as the ground truth. We execute these tests on each participant’s final code version, comparing the number of passing and failing tests between both groups. Additionally, we assessed all intermediate code versions using the commit history.

\subsubsection{RQ5: Does \toolname influence the developer experience?}

To answer this research question, participants completed an exit survey that included a series of 5-point Likert scale questions designed to assess their perceptions of the tasks and the approaches they used. Additional questions were presented to the \treatment group specifically to gauge their impressions of \toolname.

\subsubsection{RQ6: Do users of \toolname write high-quality tests?}

To address this research question, we examine the test suites created by participants using the test smell~\cite{van2001refactoring} detection system Smelly Test\footnote{\url{https://github.com/marabesi/smelly-test/}} and report the ratio of test smells to tests. \RA{Smelly Test incorporates a total of eight different test smells for TypeScript, including conditionals, timeouts, console printing, and empty tests.} In addition, we execute the tests to identify any failing ones. A failed test indicates either that the participant did not complete the test initially or that the test lacks the robustness to remain reliable over time. 

\subsubsection{RQ7: What are the most important achievements?}

To answer this research question, we analyze the final \texttt{TestReport} from each project to assess participant progress in both groups. We then calculate the mean achievement progress for each group and compute the \textit{p}-values to identify which achievements the participants aimed for.

\subsection{Threats to Validity}
\label{sec:limitations}

This subsection discusses potential validity threats in our study, following the classifications by Wohlin et al.~\cite{wohlin2012experimentation}, and our approaches to mitigate these risks.

\subsubsection{Threats to Conclusion Validity}

Threats to conclusion validity involve factors that may impair our ability to correctly infer relationships between the treatment and the study outcomes.
To address these threats, we used the Wilcoxon-Mann-Whitney test with a significance level of $\alpha=0.05$ for statistical analysis. Our data was gathered and calculated automatically with tools like Jest for test execution, which minimizes human error during data collection.
Participants were randomly assigned to one of two groups, each provided with a different operational environment. All participants were university students enrolled in two Software Engineering courses taught in different languages but covering the same English-language material. Thus, we did not consider course language a confounding factor.
While we have no evidence of varying skill levels between groups (all participants had the foundational knowledge to complete the experiment, as part of their course), we acknowledge that a different group assignment could have produced varied statistical outcomes. However, given the large sample size and knowledge distribution, this threat is unlikely, though not entirely dismissible.

\subsubsection{Threats to Internal Validity}

Threats to internal validity involve factors that may impact the causal relationship between the independent and dependent variables.
The experiment took place during regular class time, with participants using their laptops in a familiar classroom setting. While the experiment duration of 150 minutes could have led to fatigue, participants were allowed short breaks, and no one dropped out during the sessions.
Another potential threat is the unauthorized use of external tools (e.g., GitHub Copilot or ChatGPT). To mitigate this, we explicitly instructed participants not to use such tools, since task completion time and performance were not part of the evaluation, and monitored them throughout the session. However, sporadic use of these tools cannot be entirely ruled out.
Since this was a between-subjects experiment, learning biases were absent by design. Selection bias, while possible, was minimized by random group assignment and by rewarding participation rather than performance. We do not believe that the extra course grade incentive led to a disproportionate number of intrinsically motivated students.
Finally, as we used previously validated tools (i.e., \toolname, Smelly Test, Stryker, and Jest), we consider the risk of instrumentation bias negligible.

\subsubsection{Threats to Construct Validity}

Threats to construct validity concern the generalizability of the experimental results to the theoretical constructs.
%
One potential issue is that the original test suite used as ground truth may not cover all edge cases. However, since the test suite achieves 95.83\% line coverage and 100\% mutation score for the selected functions, we consider any remaining edge cases negligible for this experiment, where participants had only 150 minutes to complete and test their tasks.

\subsubsection{Threats to External Validity}

Threats to external validity refer to limitations in the generalizability of the experiment’s findings.
Our selection of participants was based on convenience, as recruiting a large student population was more feasible. Nonetheless, we argue that graduate students reasonably represent newly recruited employees~\cite{lorey2022storm}. \RB{A further iteration of this experiment can mitigate this threat by encompassing only students with working experience.}
The experimental object was a real software library, commonly used in practice, which supports our goal of generalizing the findings of the original study~\cite{original_experiment}. Our study complements previous research, expanding its findings to typical software development contexts.

	\section{Results}
	
 \label{sec:results}
 
\subsection{RQ1: Does \toolname influence testing behaviour?} \label{sec:rq1}

\paragraph{Test Creation}

\toolname aims to encourage the writing of unit tests, and it appears effective in doing so. In the \treatment group, only 10.59\% of participants did not write any Jest
tests, whereas in the \control group, this figure was higher at 21.34\%. Although this difference is close to statistical significance (exact Fisher test~\cite{bower2003use}, $p=0.064$), it is not fully conclusive. Additionally, participants in the \control group relied more often on the main method for manual testing: 73.03\% of the \control group used the `main` method at some point for testing, compared to only 55.29\% in the \treatment group, which is a significant difference ($p=0.018$).

\begin{figure*}
		\centering
		\begin{subfigure}[t]{0.3\textwidth}
			\centering
			\includegraphics[width=\textwidth]{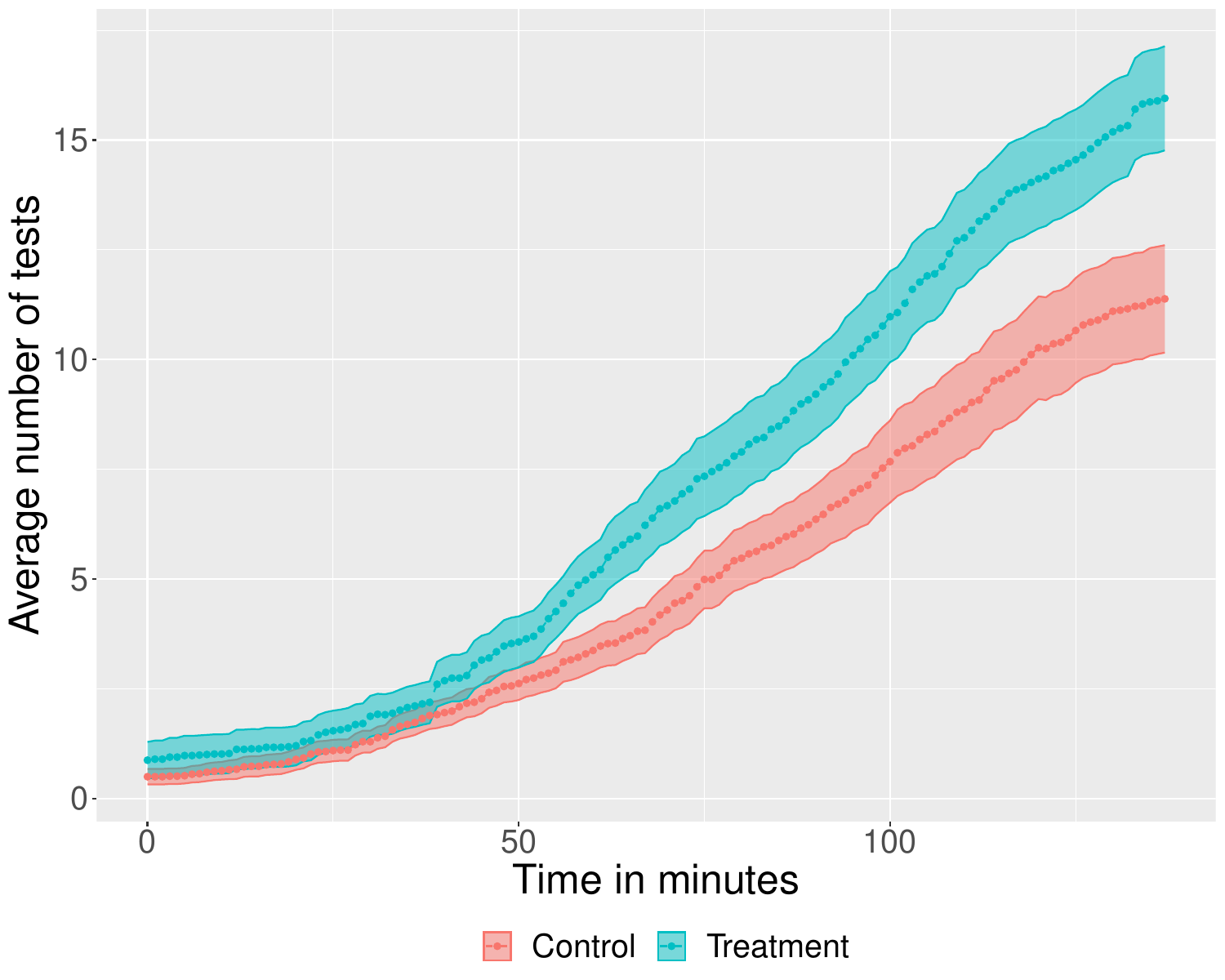}
			\vspace{-1em}
			\caption{Number of tests created over time}
			\label{fig:teststime}
		\end{subfigure}
		\hfill
		\begin{subfigure}[t]{0.3\textwidth}
			\centering
			\includegraphics[width=\textwidth]{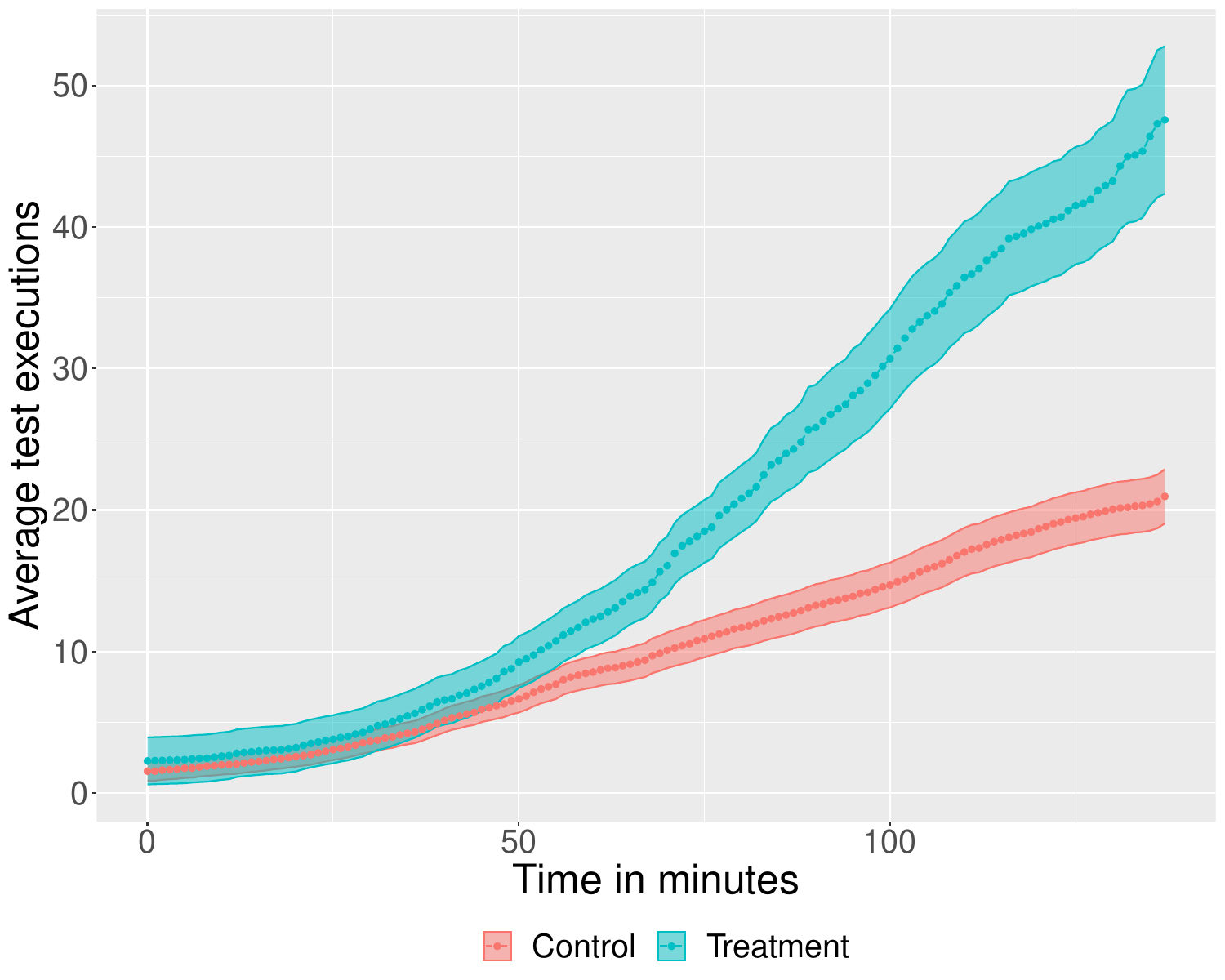}
			\vspace{-1em}
			\caption{Number of test executions over time}
			\label{fig:testexecutionstime}
		\end{subfigure}
		\hfill
		\begin{subfigure}[t]{0.3\textwidth}
			\centering
			\includegraphics[width=\textwidth]{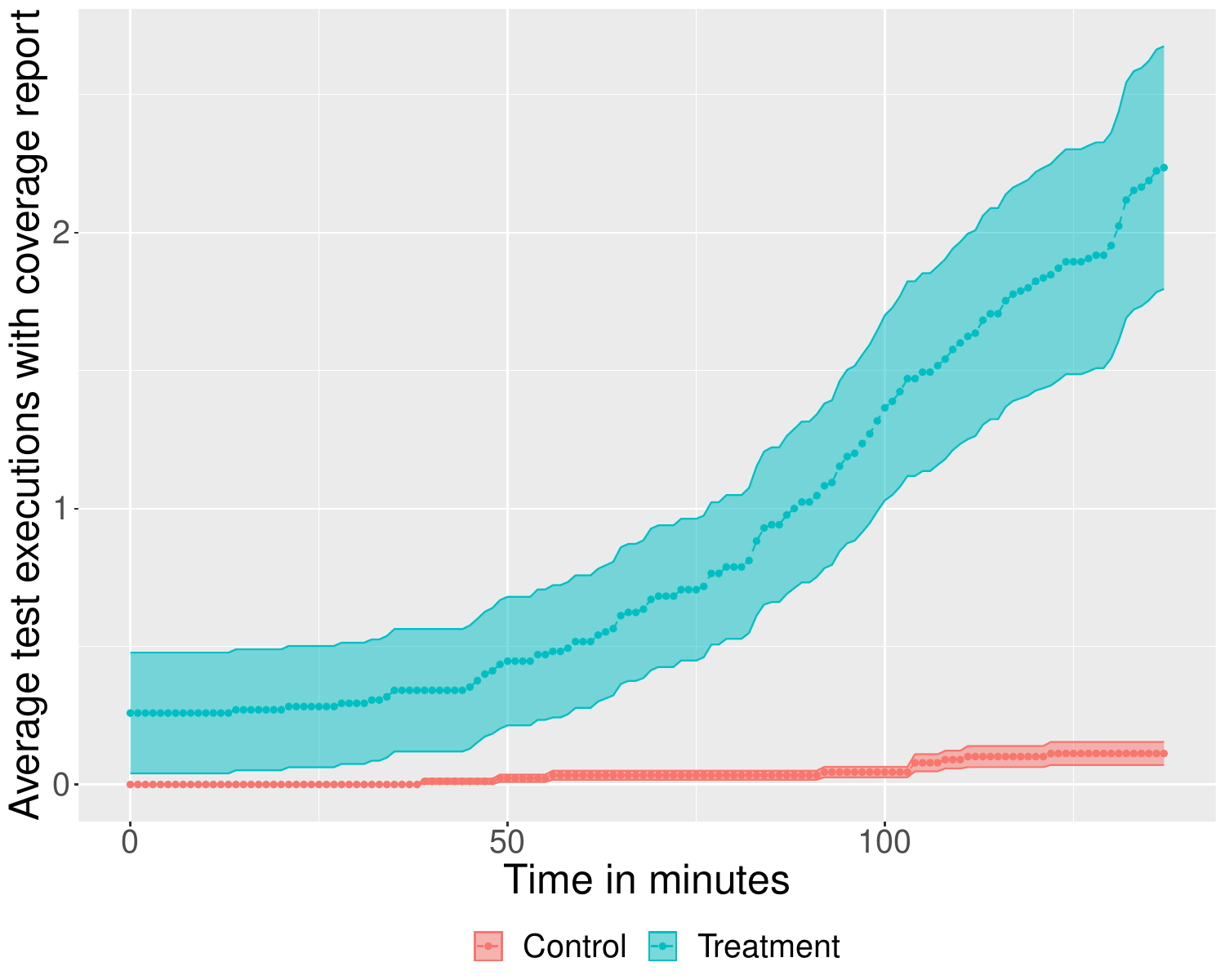}
			\vspace{-1em}
			\caption{Number of test executions with coverage over time}
			\label{fig:testexecutionscoveragetime}
		\end{subfigure}
		\hfill
		\begin{subfigure}[t]{0.3\textwidth}
			\centering
			\includegraphics[width=\textwidth]{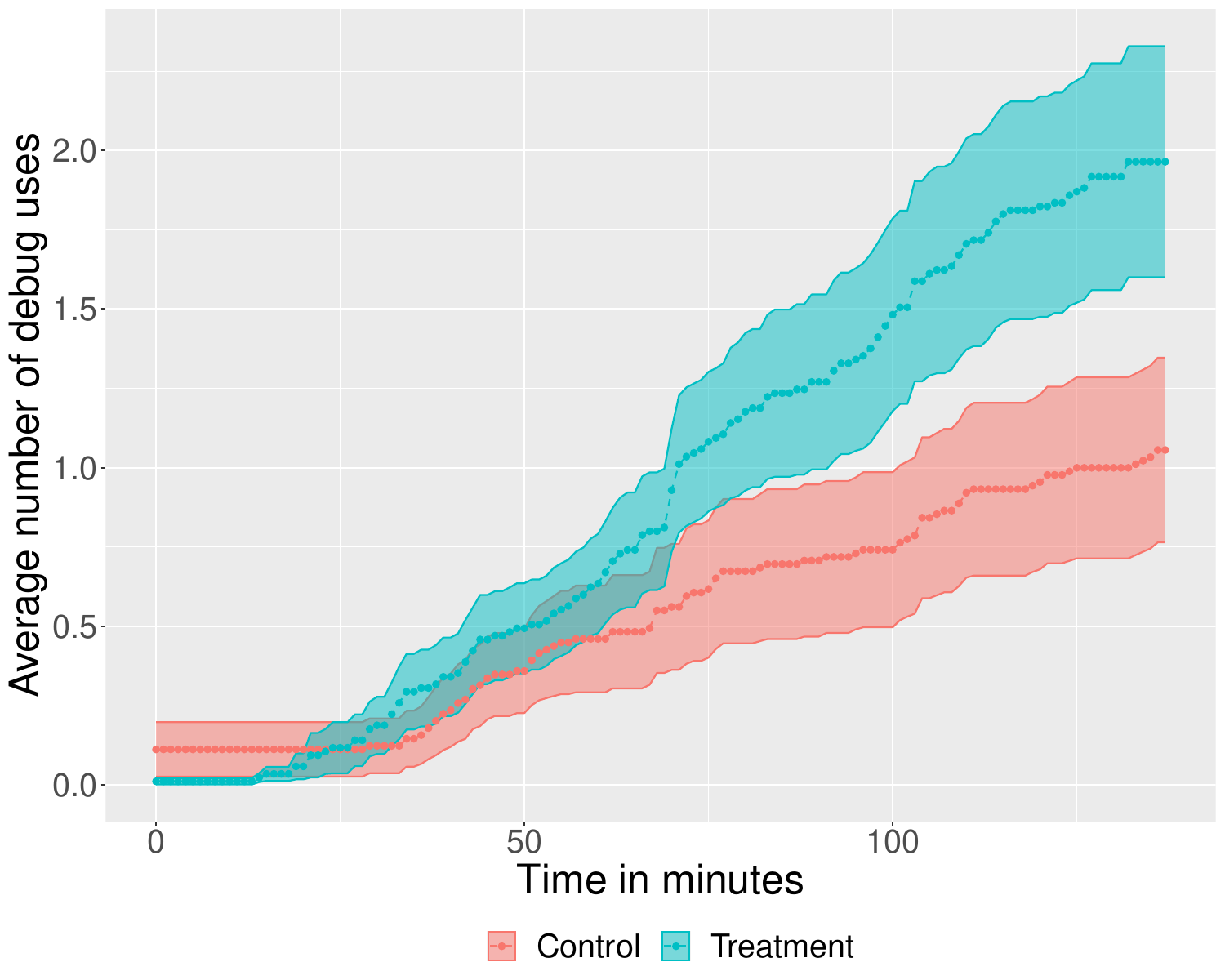}
			\vspace{-1em}
			\caption{Number of debug uses over time}
			\label{fig:debugtime}
		\end{subfigure}
		\hfill
		\begin{subfigure}[t]{0.3\textwidth}
			\centering
			\includegraphics[width=\textwidth]{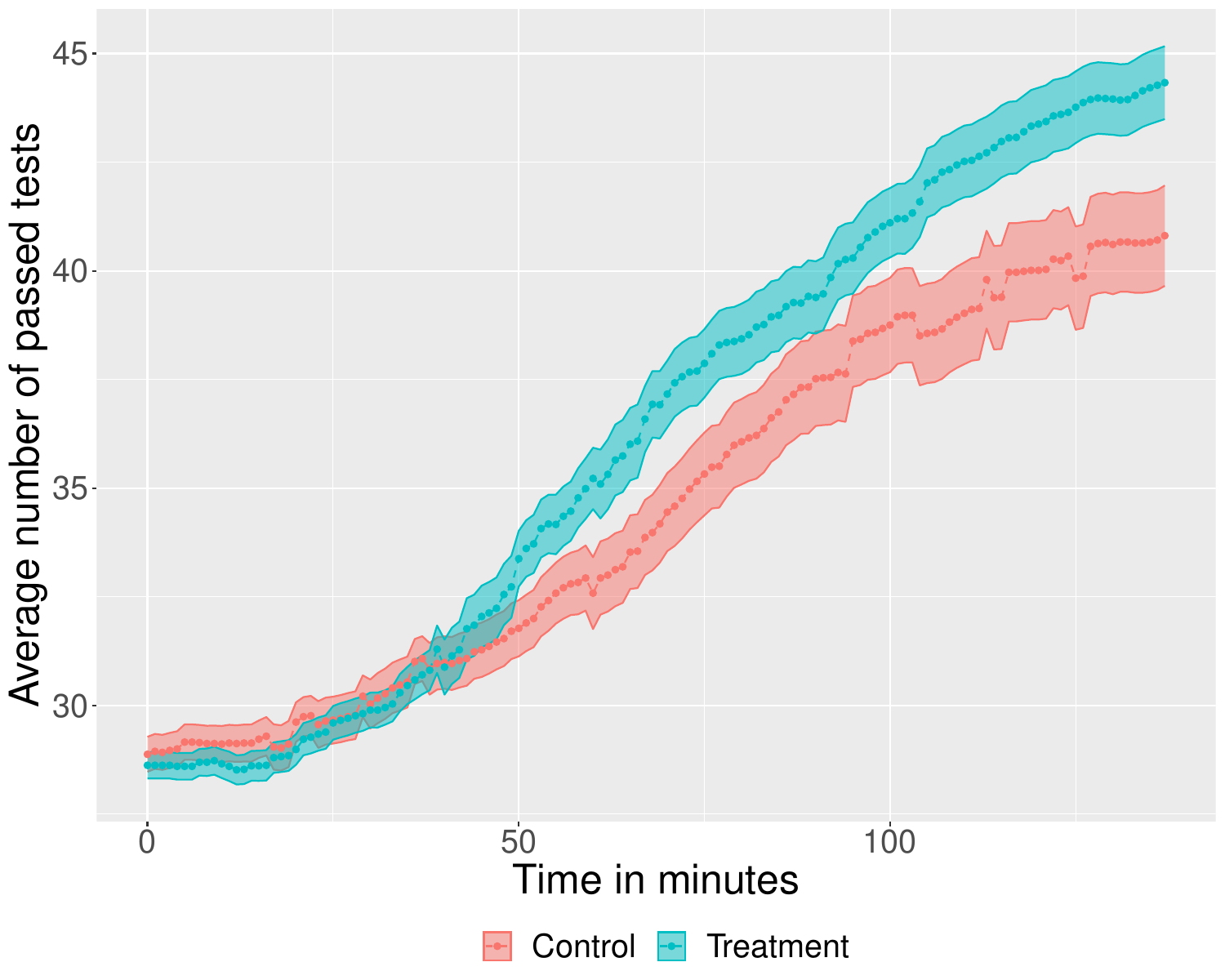}
			\vspace{-1em}
			\caption{Number of passed tests over time}
			\label{fig:passedteststime}
		\end{subfigure}
		\hfill
		\begin{subfigure}[t]{0.3\textwidth}
			\centering
			\includegraphics[width=\textwidth]{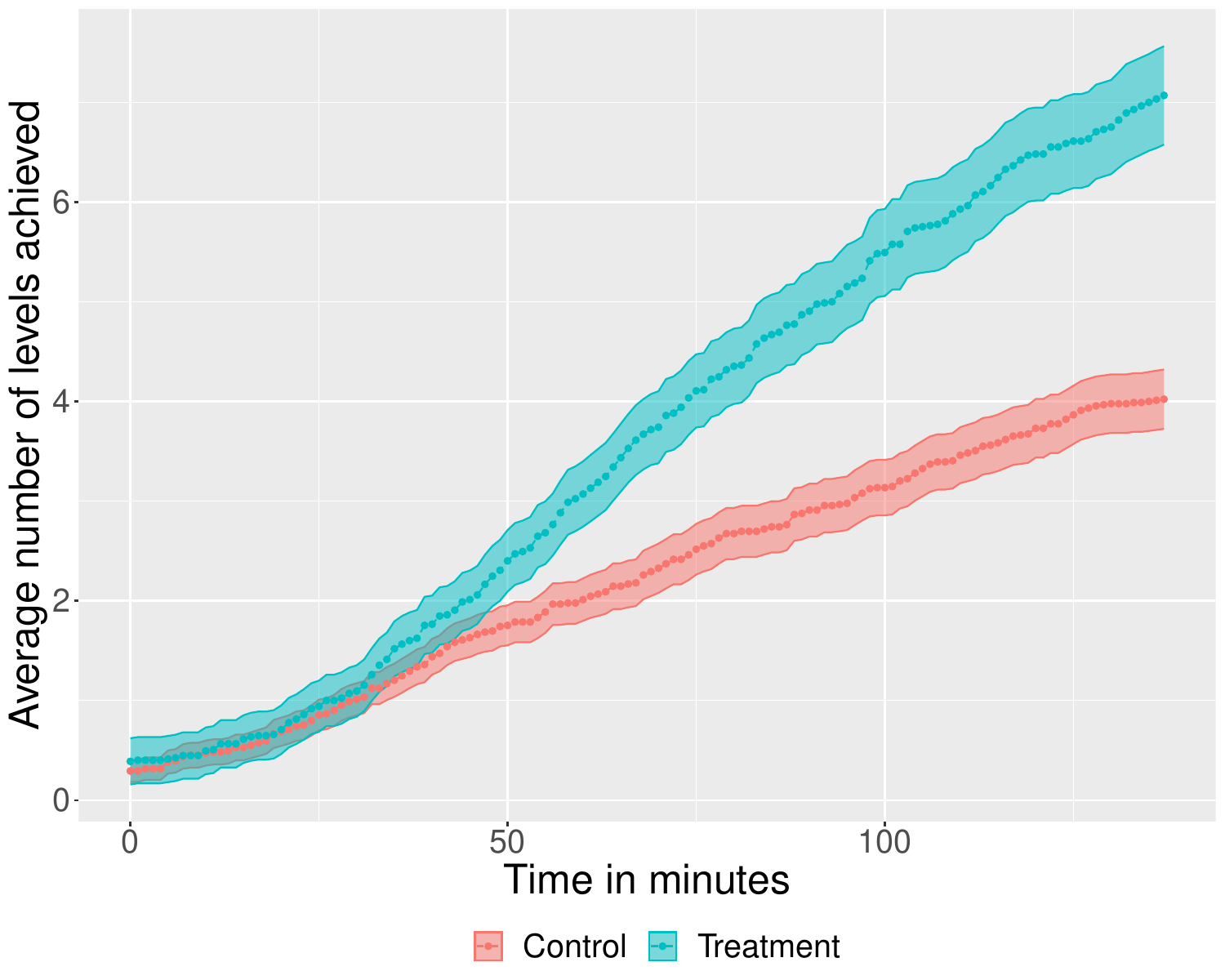}
			\vspace{-1em}
			\caption{Number of levels over time}
			\label{fig:levelstime}
		\end{subfigure}
		
		\caption{Differences between \control and \treatment groups over time}
		\label{fig:timeplots}
	\end{figure*}

\Cref{fig:teststime} illustrates the progression of Jest tests written over time. Both the \treatment and \control groups began writing tests early on, right from the start. However, the \treatment group consistently wrote more tests from the outset and showed a sharper increase in test writing compared to the \control group, starting around minute 39. From minute 53 until the end, the difference between the groups is statistically significant, as indicated by non-overlapping confidence intervals.

\paragraph{Test Executions}

\begin{figure*}
    \centering
    \begin{subfigure}[t]{0.245\textwidth}
        \centering
        \includegraphics[width=\textwidth]{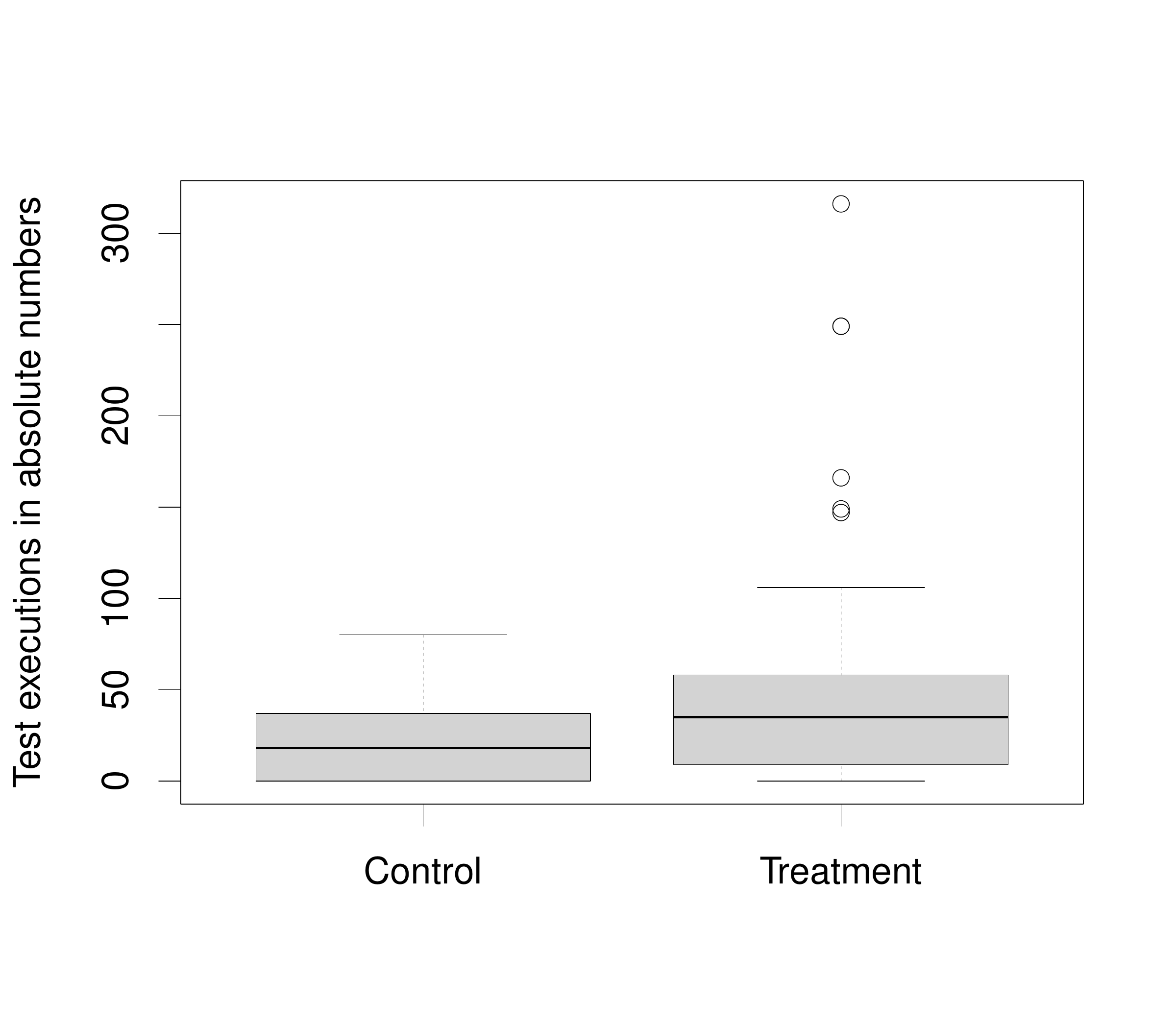}
        \vspace{-2.5em}
        \captionsetup{justification=centering}
        \caption{Number of test executions during the experiment}
        \label{fig:testexecutions}
    \end{subfigure}
    \hfill
    \begin{subfigure}[t]{0.245\textwidth}
        \centering
        \includegraphics[width=\textwidth]{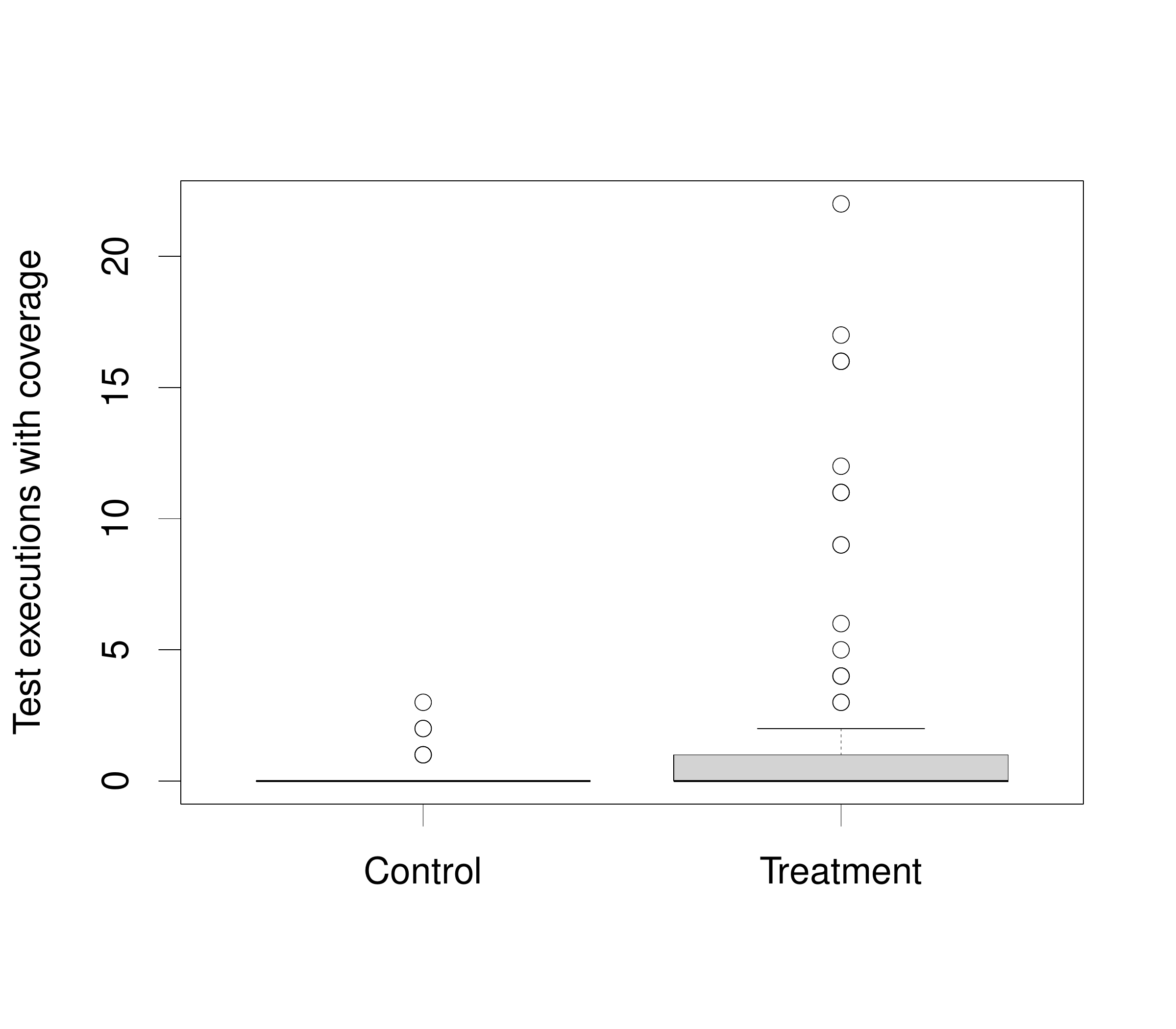}
        \vspace{-2.5em}
        \captionsetup{justification=centering}
        \caption{Number of test executions with coverage in IntelliJ}
        \label{fig:testexecutionscoverage}
    \end{subfigure}
    \hfill
    \begin{subfigure}[t]{0.245\textwidth}
        \centering
        \includegraphics[width=\textwidth]{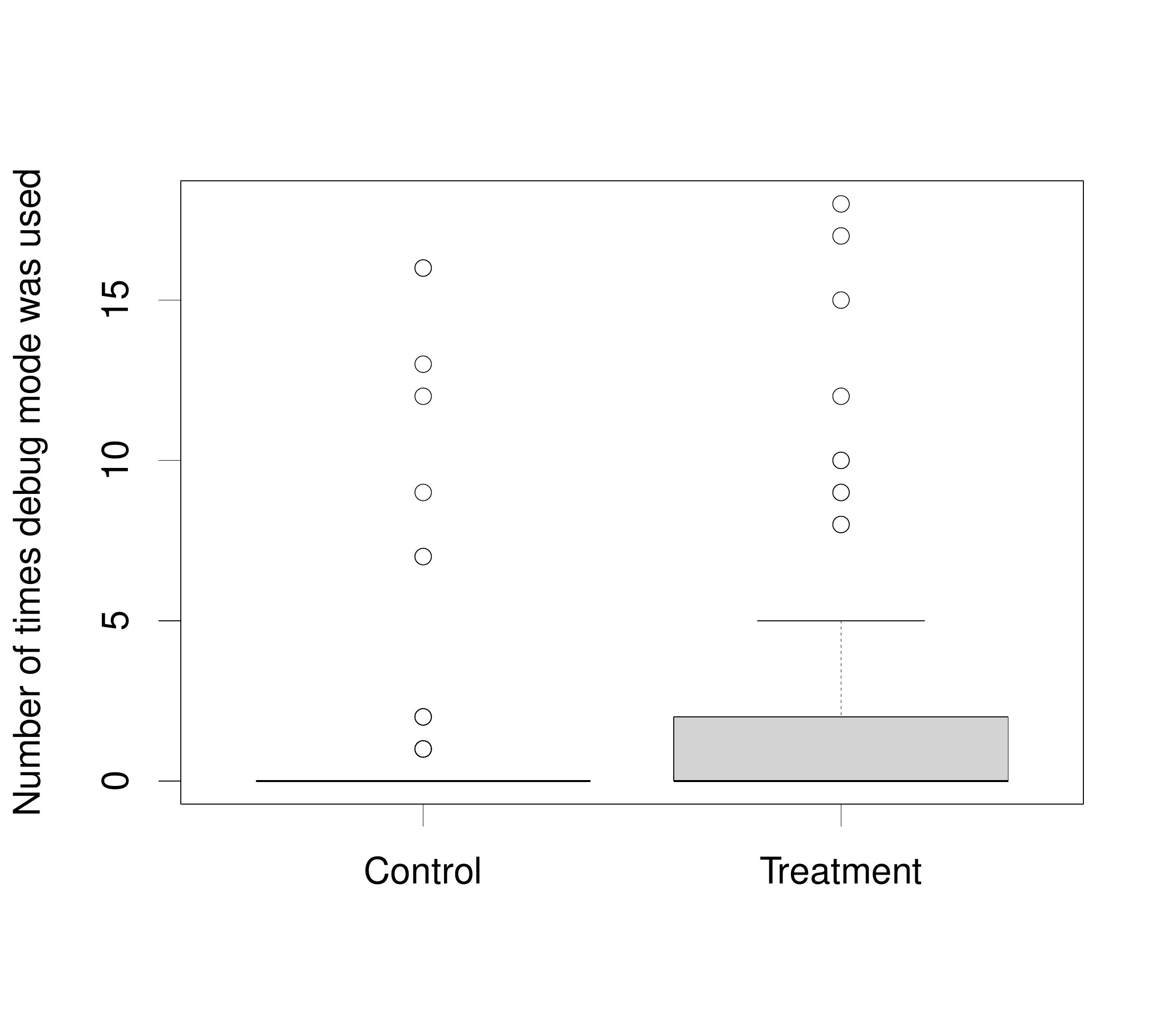}
        \vspace{-2.5em}
        \captionsetup{justification=centering}
        \caption{Number of times the IntelliJ debug mode was used}
        \label{fig:debugmode}
    \end{subfigure}
    \hfill
    \begin{subfigure}[t]{0.245\textwidth}
        \centering
        \includegraphics[width=\textwidth]{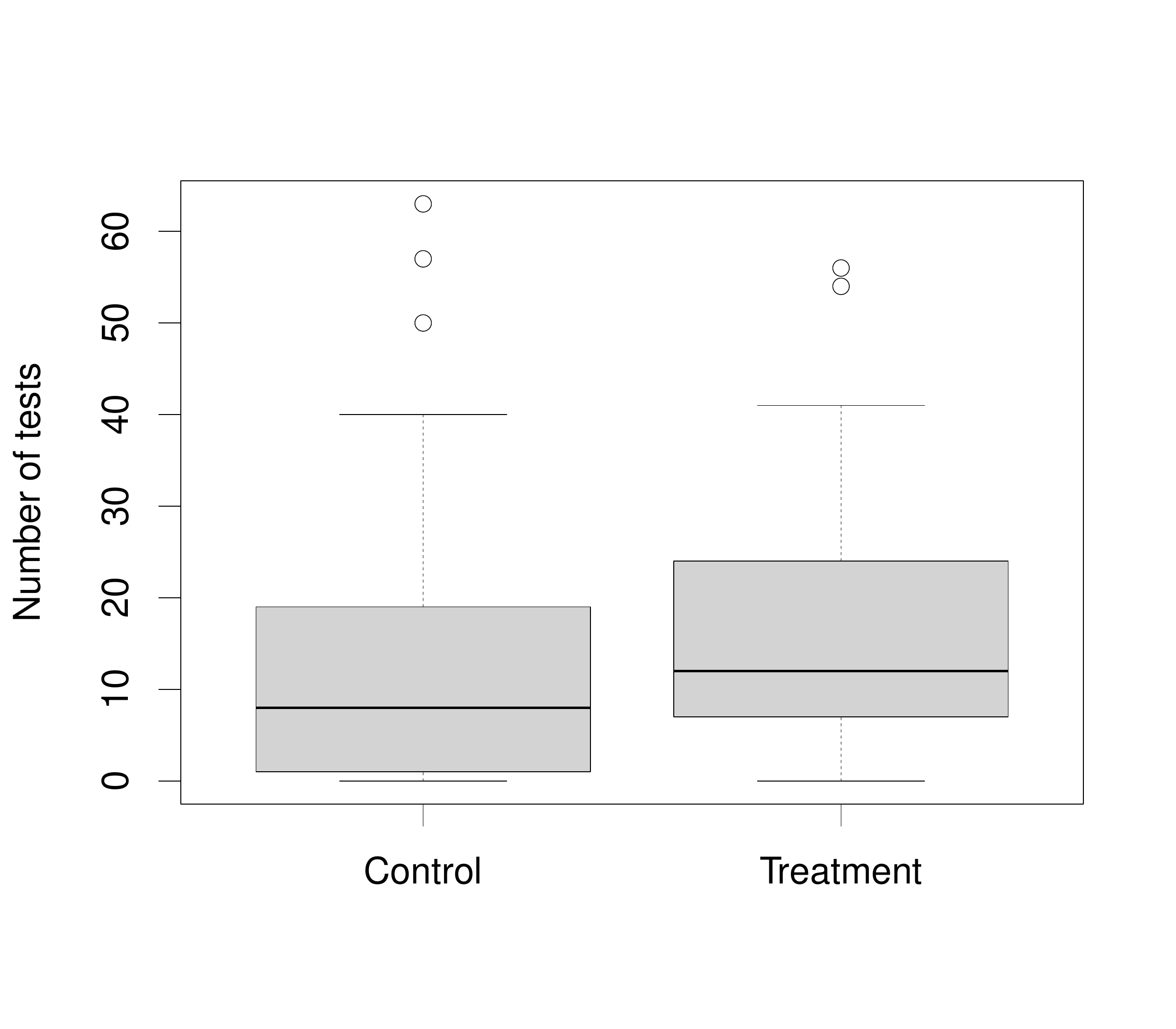}
        \vspace{-2.5em}
        \captionsetup{justification=centering}
        \caption{Number of tests written by the participants}
        \label{fig:testnumber}
    \end{subfigure}
    \hfill
    \begin{subfigure}[t]{0.245\textwidth}
        \centering
        \includegraphics[width=\textwidth]{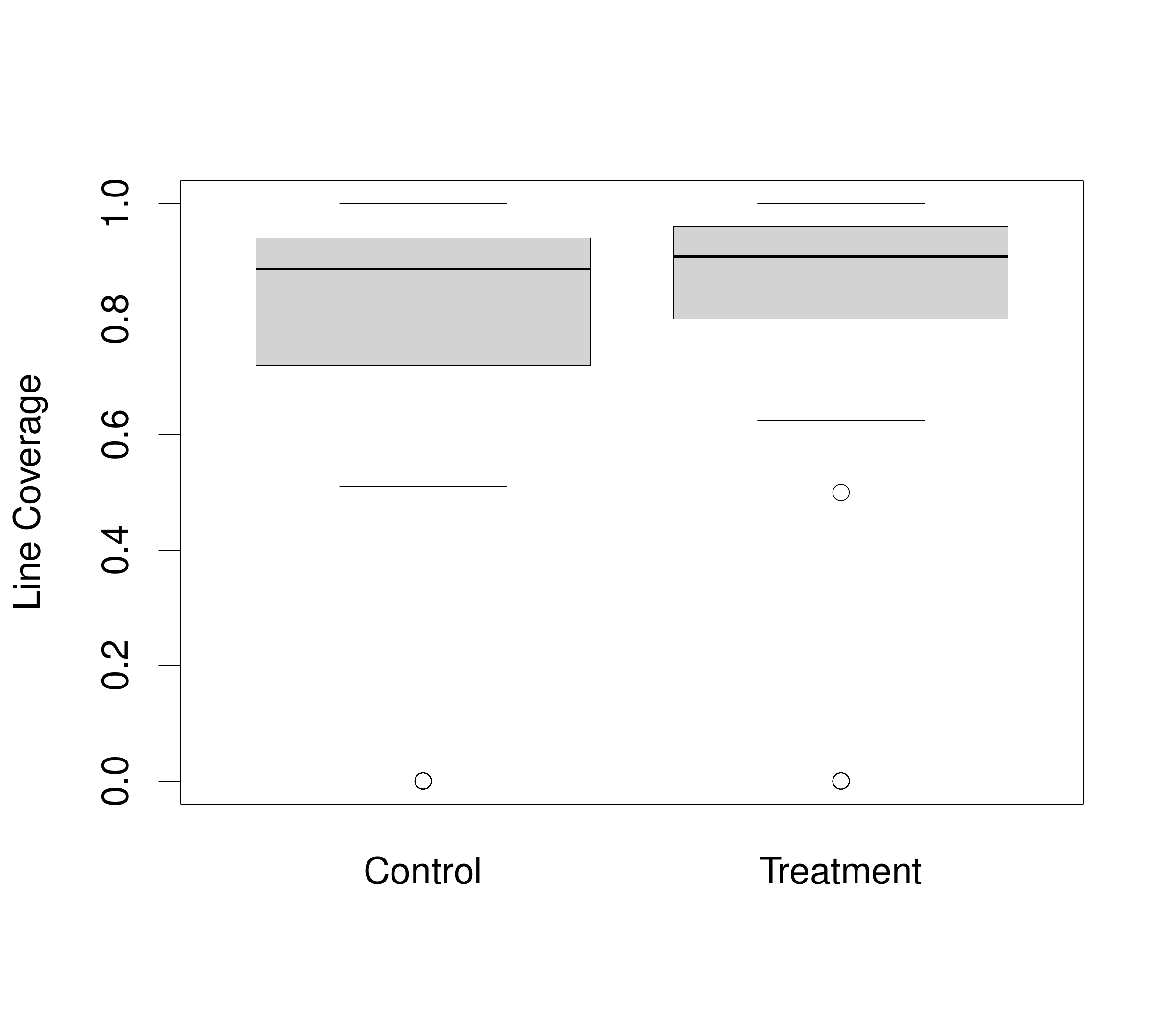}
        \vspace{-2.5em}
        \captionsetup{justification=centering}
        \caption{Line coverage of the final test suites}
        \label{fig:linecoverage}
    \end{subfigure}
    \hfill
    \begin{subfigure}[t]{0.245\textwidth}
        \centering
        \includegraphics[width=\textwidth]{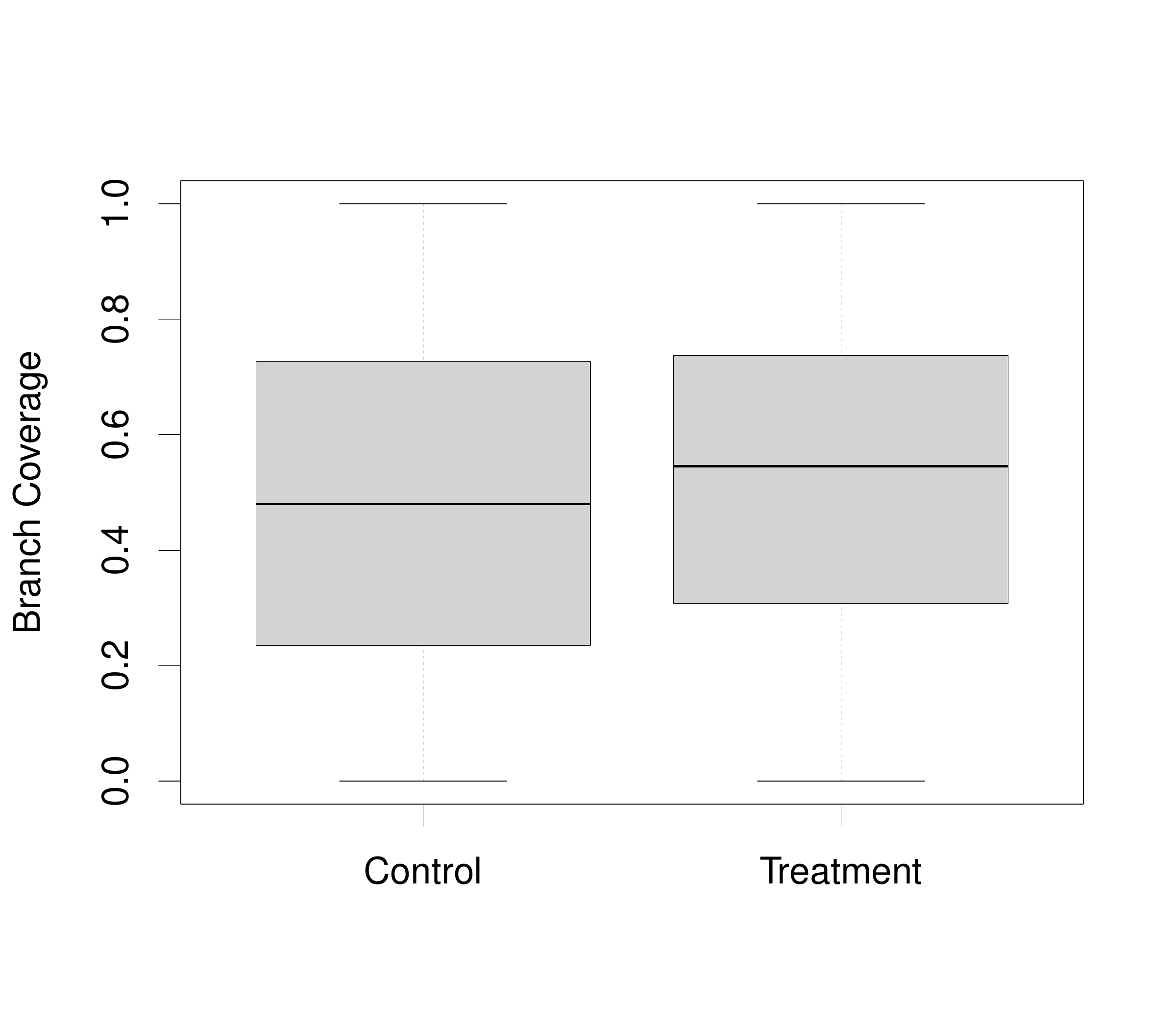}
        \vspace{-2.5em}
        \captionsetup{justification=centering}
        \caption{Branch coverage of the final test suites}
        \label{fig:branchcoverage}
    \end{subfigure}
    \hfill
    \begin{subfigure}[t]{0.245\textwidth}
        \centering
        \includegraphics[width=\textwidth]{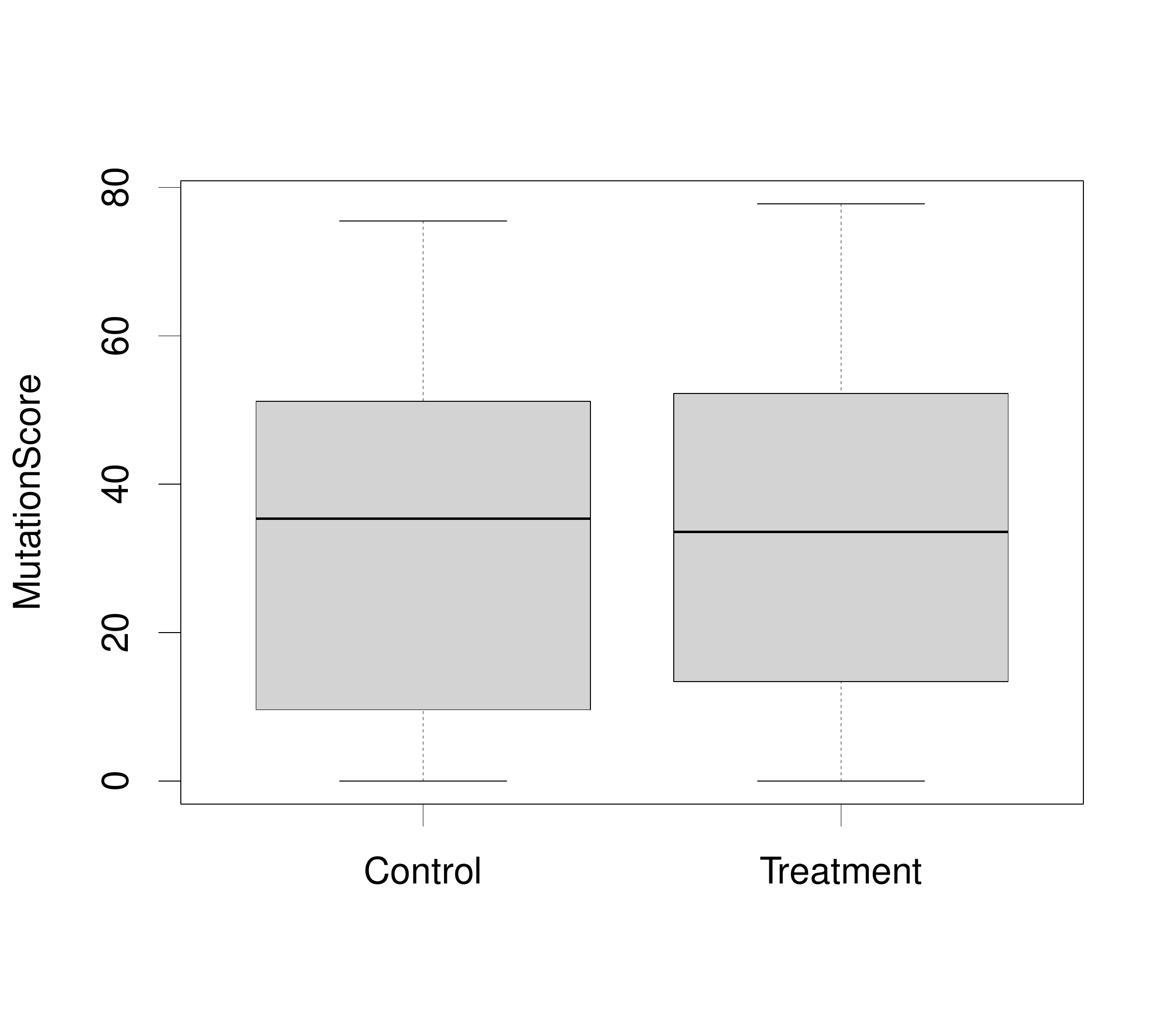}
        \vspace{-2.5em}
        \captionsetup{justification=centering}
        \caption{Mutation scores of the final test suites}
        \label{fig:mutationscore}
    \end{subfigure}
    \hfill
    \begin{subfigure}[t]{0.245\textwidth}
        \centering
        \includegraphics[width=\textwidth]{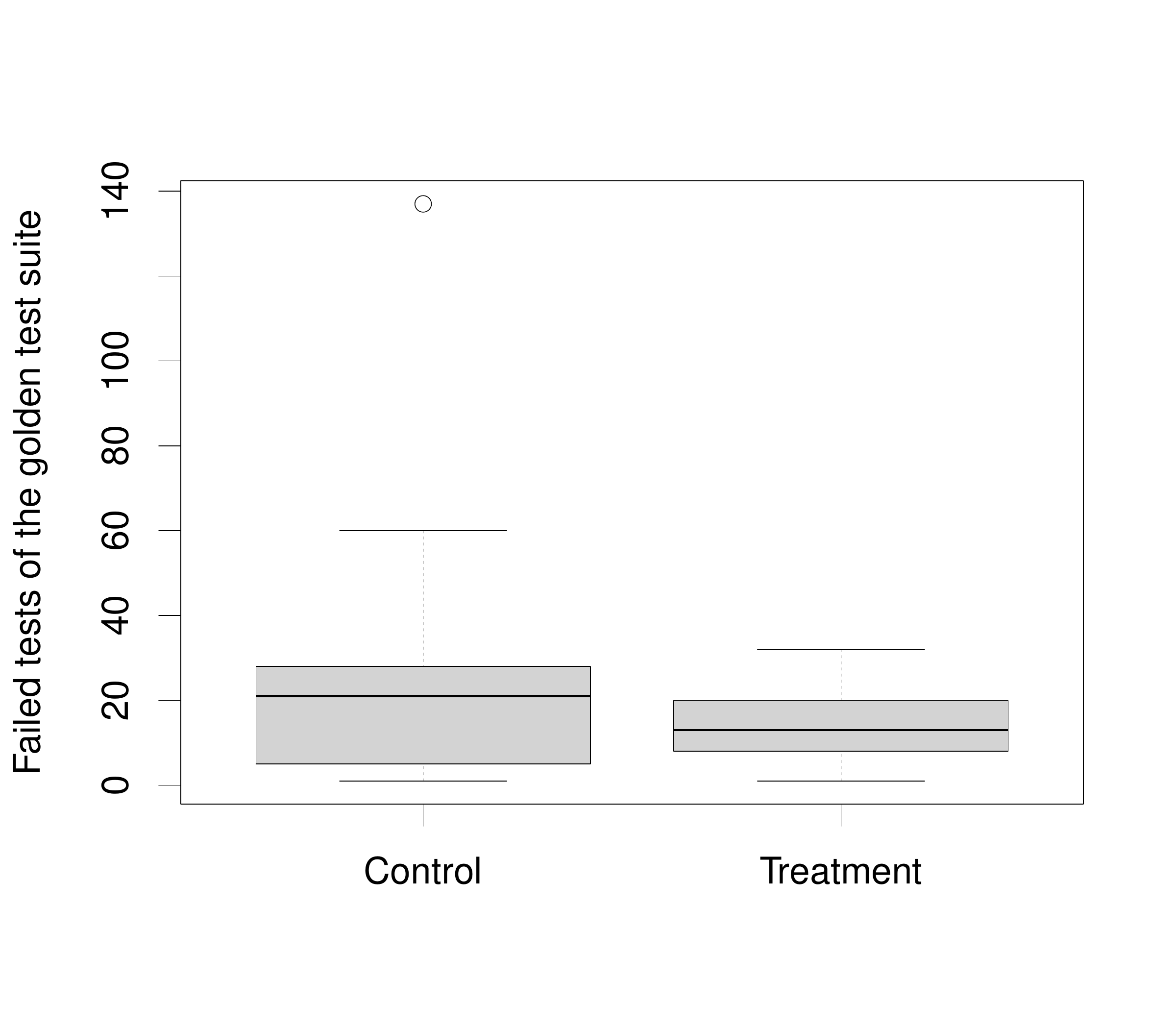}
        \vspace{-2.5em}
        \captionsetup{justification=centering}
        \caption{Number of failed tests according to the golden test suite}
        \label{fig:testsfailed}
    \end{subfigure}
    
    \caption{Differences between \control and \treatment groups}
    \label{fig:boxplots}
\end{figure*}

\Cref{fig:testexecutions} displays the number of test executions, with an average of 47.56 in the \treatment group and 20.97 in the \control group—a significant difference ($p<0.001$). The \treatment group also had a notably higher maximum number of executions, reaching 316 compared to just 80 in the \control group.  Additionally, \cref{fig:testexecutionstime} shows test executions over time, along with the 84.6\% confidence intervals. A significant difference between both groups emerges at minute 56, indicating that the achievements in \toolname motivated participants to execute their tests more frequently.

\paragraph{Coverage Measurement}

\Cref{fig:testexecutionscoverage} compares test executions with coverage collection enabled. In the \control group, only 6.74\% of participants used this IntelliJ feature, resulting in an overall average of 0.11 coverage executions per participant. In contrast, the \treatment group used the coverage report more frequently, with an average of 2.24 executions per participant. \Cref{fig:testexecutionscoveragetime} shows coverage executions over time, revealing that the first participant in the \treatment group used the coverage report from the very start, while the \control group only began using it after 39 minutes. This indicates a significant difference ($p<0.001$) from the beginning, as the \treatment group participants demonstrated greater engagement with coverage collection.

\paragraph{Debug Executions}

\Cref{fig:debugmode} shows the frequency of test or program executions in debug mode. Participants in the \treatment group used debug mode an average of 1.96 times, compared to 1.06 times in the \control group—a significant difference ($p=0.015$). This difference is also evident in \cref{fig:debugtime}, where confidence intervals diverge after minute 79. Some control group participants initially used debug mode instead of regular runs, explaining their early lead. Given the program's complexity, particularly in date and time handling, participants increasingly relied on debug mode in later stages, with achievements in \toolname likely encouraging more frequent use.

\summary{RQ 1}{\toolname significantly influences participants' testing behavior: they prefer using Jest tests over main testing, create more tests, run their tests more frequently, make greater use of coverage reports, and rely more often on debug mode for both program and test debugging.}

\comparison{Our findings confirm the results of RQ1 from the original paper and additionally show that \toolname also affects debugging behavior.}

\subsection{RQ2: \RA{How} does \toolname influence resulting test suites?} \label{sec:rq2}

\paragraph{Number of Tests}

\Cref{fig:testnumber} compares the final test suites based on the number of Jest tests they contain. Participants in the \treatment group wrote an average of 15.95 tests, whereas the \control group averaged 11.38 tests. The exact Wilcoxon-Mann-Whitney test confirms that the \treatment group wrote significantly more tests than the \control group ($p=0.003$).
\RB{
We found indication that they focused on more fine-grained testing, for example by using Boundary Value Analysis (BVA)~\cite{DBLP:conf/euromicro/Ramachandran03} to test dates just inside or outside valid ranges. However, to some extent the increased test count may also simply be due to \toolname encouraging more tests, possibly including some redundant ones. }

\paragraph{Code Coverage}

\Cref{fig:linecoverage} compares line coverage between the \treatment and \control groups, showing an average coverage of 81\% and 75\%, respectively. Although the \treatment group thus achieved higher coverage, this difference is not statistically significant ($p=0.14$). In both groups, some participants achieved 0\% line coverage, indicating they did not write any unit tests and relied on the main method for testing. Similar results are observed for branch coverage (\cref{fig:branchcoverage}), with average coverages of 51\% in the \treatment group and 46\% in the \control group ($p=0.34$). Overall, although the \treatment group had slightly higher average coverage, the difference was not statistically significant. Given that participants received documentation of the methods to implement, achieving coverage was relatively straightforward, which may account for the lack of a significant difference in coverage compared to the original study.

\paragraph{Mutation Score}

The mean mutation score was 34.08\% for the \treatment group and 32.4\% for the \control group, a non-significant difference ($p=0.58$) (\cref{fig:mutationscore}). This suggests that both groups approached test writing similarly, leading to comparable coverage and mutation scores. The mutation analysis excluded failing tests, which may explain the lack of significant difference in mutation scores. \RBremoved{Given the project’s focus on date and time functionality, these timing differences could influence mutation scores. If all tests had passed, there might have been a more pronounced difference in mutation scores between the groups.} \RB{Upon investigating failing tests, we observed that many of them are related to date discrepancies, as we reran all tests for analysis in October 2024, about four months after the experiment. We will investigate these failing tests and test quality further in \cref{sec:rq6}. However, to understand whether the lack of significant improvement of the mutation score is a result of the date difference in our analysis, we re-ran the mutation analysis with the system time being set to the time and date of the commit, and while the mutation scores increase for both groups, the difference remains non-significant.}

\summary{RQ 2}{\toolname led the \treatment group to write significantly more Jest tests than the \control group. \RB{However, since code coverage and mutation scores remained similar between the groups, this suggests that the tool may have encouraged quantity over effectiveness in test writing.}}

\comparison{Our findings confirm that \toolname influences test suite outcomes by increasing the number of tests. However, we did not observe significant differences in code coverage or mutation scores, so we were unable to replicate the original study’s results in these areas.}

\begin{figure}
		\centering
		\begin{subfigure}[b]{0.24\linewidth}
			\centering
			\includegraphics[width=\linewidth]{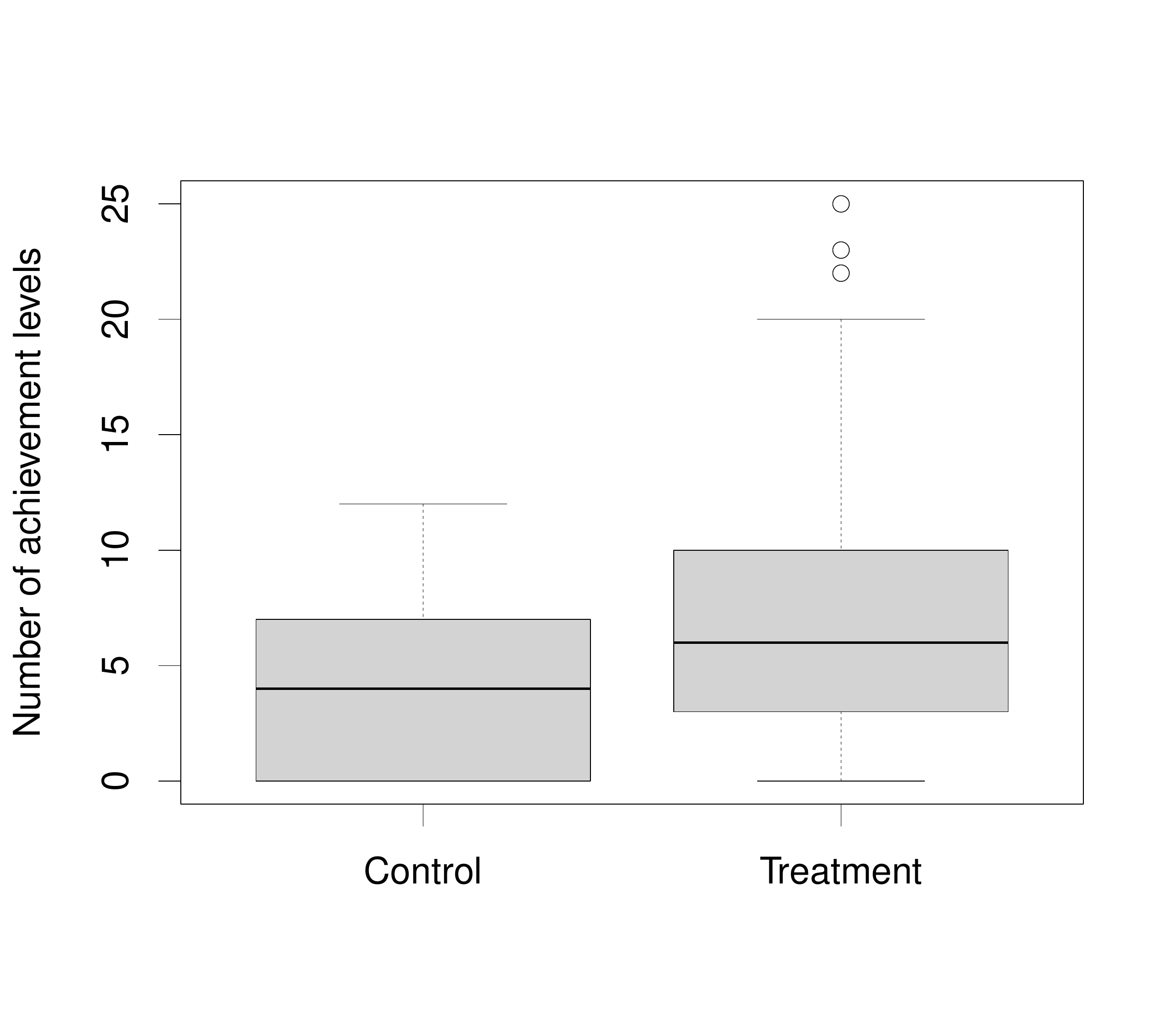}
			\caption{The number of achievement levels reached by the participants}
			\label{fig:levels}
		\end{subfigure}
		\hfill
		\begin{subtable}[b]{0.24\linewidth}
			\tiny
			\begin{tabularx}{\linewidth}{lll}
				\toprule
				\textbf{Variable} & $\mathbf{r^2}$ & \textbf{p-value}   \\ \midrule
				Line Coverage & 0.42 & < 0.001 \\
				Branch Coverage & 0.38 & < 0.001 \\
				Mutation Score & 0.48 & < 0.001 \\
				Number of Tests & 0.65 & < 0.001 \\
				\bottomrule
			\end{tabularx}%
			\setlength{\tabcolsep}{6pt}
                \vspace{0.65cm}
			\caption{The Pearson correlations based on the number of levels}
			\label{tab:corr}
		\end{subtable}
            \hfill
            \begin{subfigure}[b]{0.24\linewidth}
			\centering
			\includegraphics[width=\linewidth]{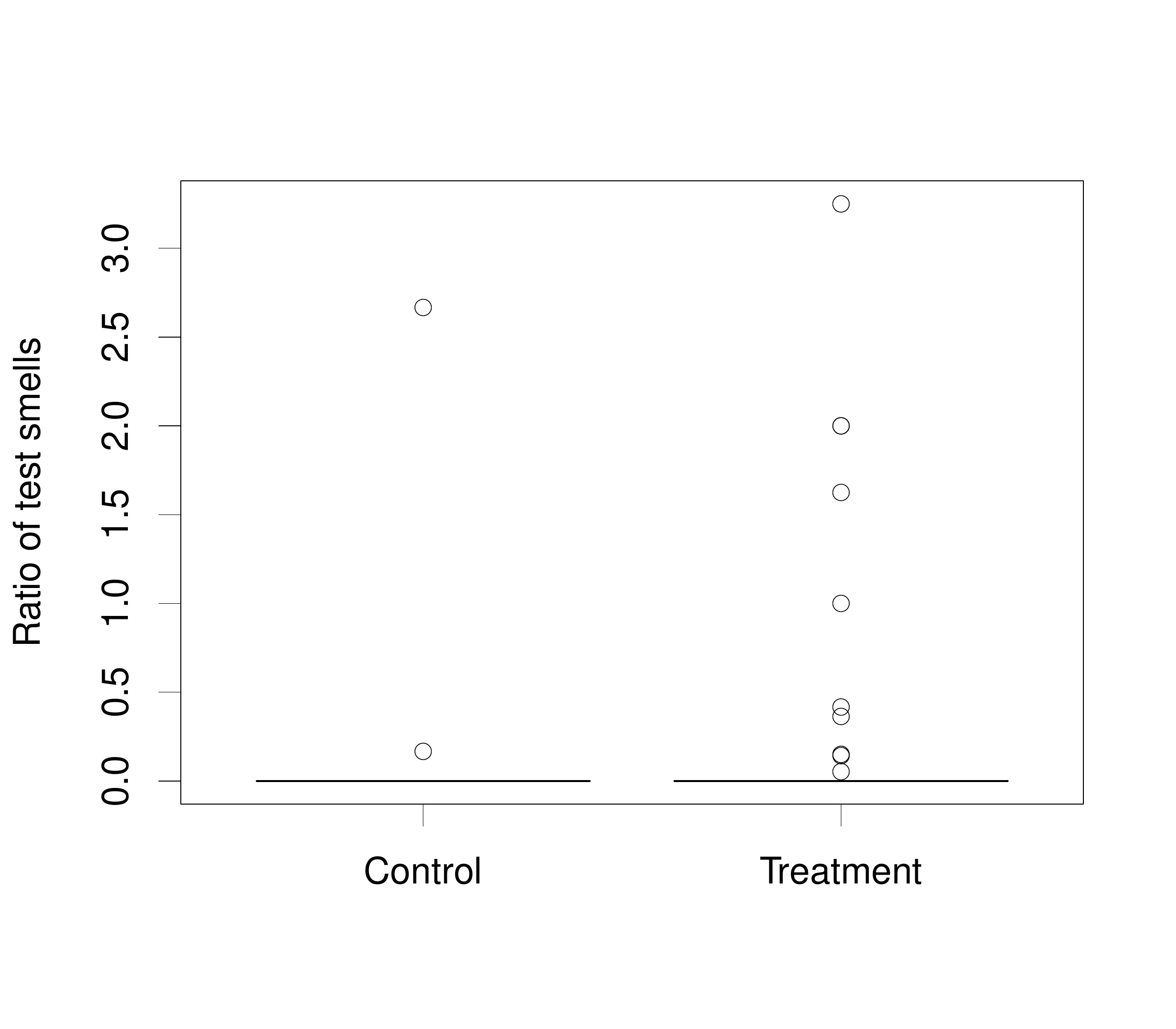}
			\caption{The ratio of test smells in the tests of the participants}
			\label{fig:smells}
		\end{subfigure}
            \hfill
            \begin{subfigure}[b]{0.24\linewidth}
			\centering
			\includegraphics[width=\linewidth]{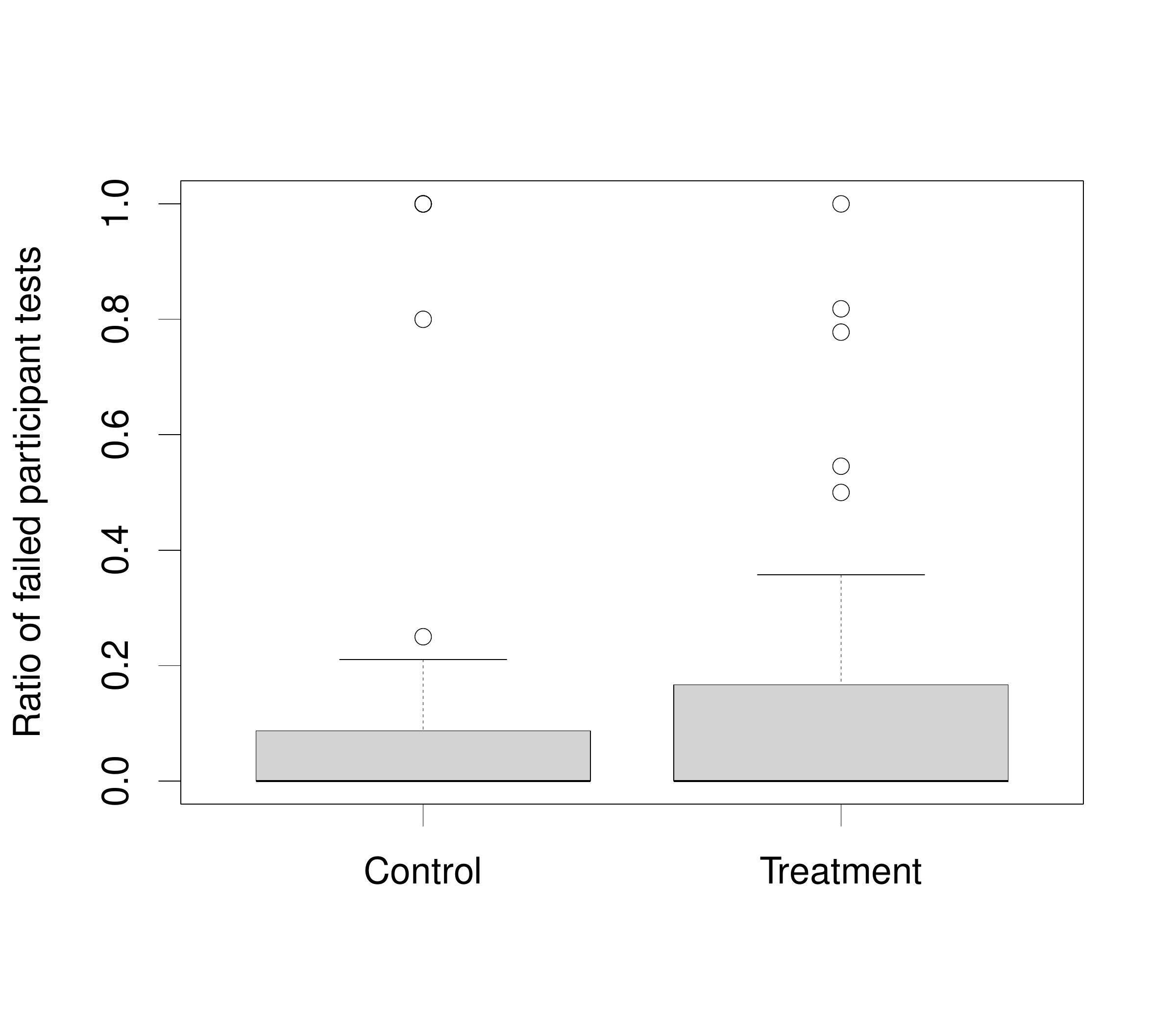}
			\caption{The ratio of failed tests written by the partici-\\pants}
			\label{fig:failed}
		\end{subfigure}
		
		\caption{The number of achievement levels, their Pearson correlations with different test suite metrics as well as the ratios of failed tests and test smells}
		\label{fig:levcorr}
	\end{figure}

\subsection{RQ3: Do achievement levels reflect differences in test suites and activities?}

RQ1 demonstrated that participants in the \treatment group are more actively engaged in testing activities compared to those in the \control group. \Cref{fig:levels} compares the achievement levels reached by participants in both groups, showing a mean of 4.02 for the \control group and 7.07 for the \treatment group. \Cref{fig:levelstime} indicates that this difference becomes evident starting at minute 48 of the experiment. The difference is significant, as confirmed by the exact Wilcoxon-Mann-Whitney test $p<0.001$. Thus, developers who earn achievements are indeed more committed to testing.

To assess whether developers with higher achievement levels produce better test suites overall, \cref{tab:corr} presents the Pearson rank correlations between test suite metrics (RQ2) and achievement levels. A strong significant correlation exists between the number of tests and achievement levels, along with a moderate positive correlation for both line coverage and mutation score. Additionally, there is a weak significant correlation between achievement levels and branch coverage. These findings strongly support the effectiveness of the gamification approach, demonstrating that earning achievements is associated with producing better test suites.

\summary{RQ 3}{Higher achievement levels indicate greater motivation for testing and lead to better-quality test suites.}

\comparison{Our findings support those of the original paper, showing that higher achievement levels lead to increased motivation and improved test suites.}

\subsection{RQ4: Does \toolname influence the functionality of the resulting code?}

\Cref{fig:passedteststime} illustrates the number of passed tests in the golden test suite throughout the experiment, as a proxy for how much of the target functionality is already correctly implemented. Notably, the \treatment group began passing tests earlier, starting at minute 50, compared to the \control group. Although there is a small overlap around minute 90, the differences remain significant until the end of the experiment since the graphs do not overlap anymore. Since some implementations were provided, all participants had at least 28 passing tests out of the 60 in the golden test suite.

By the end of the experiment, the average number of failing tests in the golden test suite (\cref{fig:testsfailed}) was 14.32 for the \treatment group and 19.25 for the \control group, showing a nearly statistically significant difference ($p=0.054$). This indicates that participants in the \treatment group implemented more functionality correctly by the end of the experiment, while the \control group had a higher number of errors in their implementations.

\summary{RQ 4}{Users of \toolname demonstrated earlier passing and fewer failing tests in the golden test suite, indicating better implementation of functionality compared to the \control group.}

\comparison{Our findings confirm that \toolname encourages earlier implementation of functionality and, unlike the original study, also shows that \toolname enhances functionality by the end of the experiment.}

\begin{figure}
    \begin{subfigure}[t]{0.49\textwidth}
        \includegraphics[width=\textwidth]{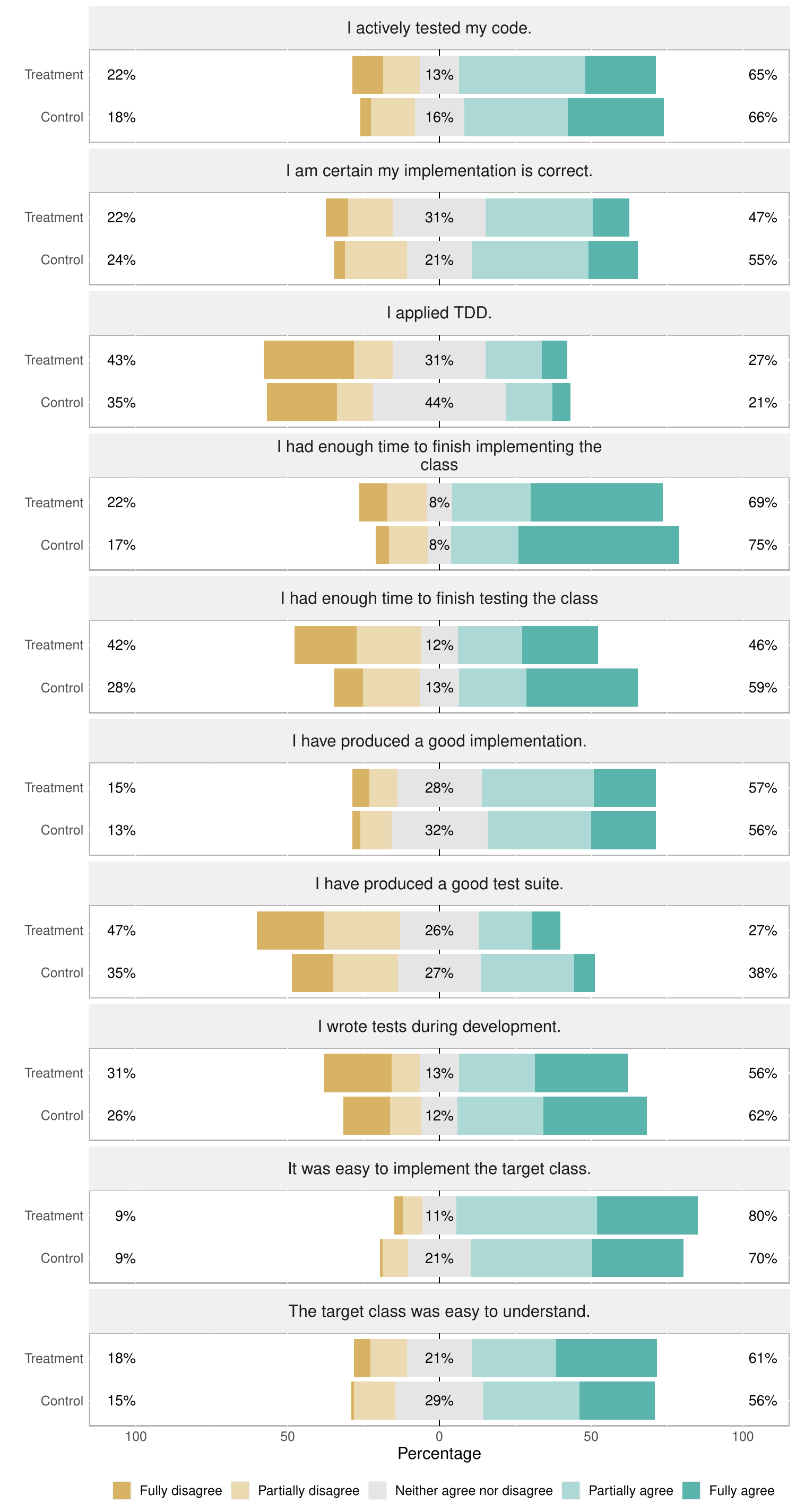}
        \caption{General survey responses---ranging from negative on the left to positive on the right}
        \label{fig:exitsurveyall}
    \end{subfigure}
    \hfill
    \begin{subfigure}[t]{0.49\textwidth}
        \includegraphics[width=\textwidth]{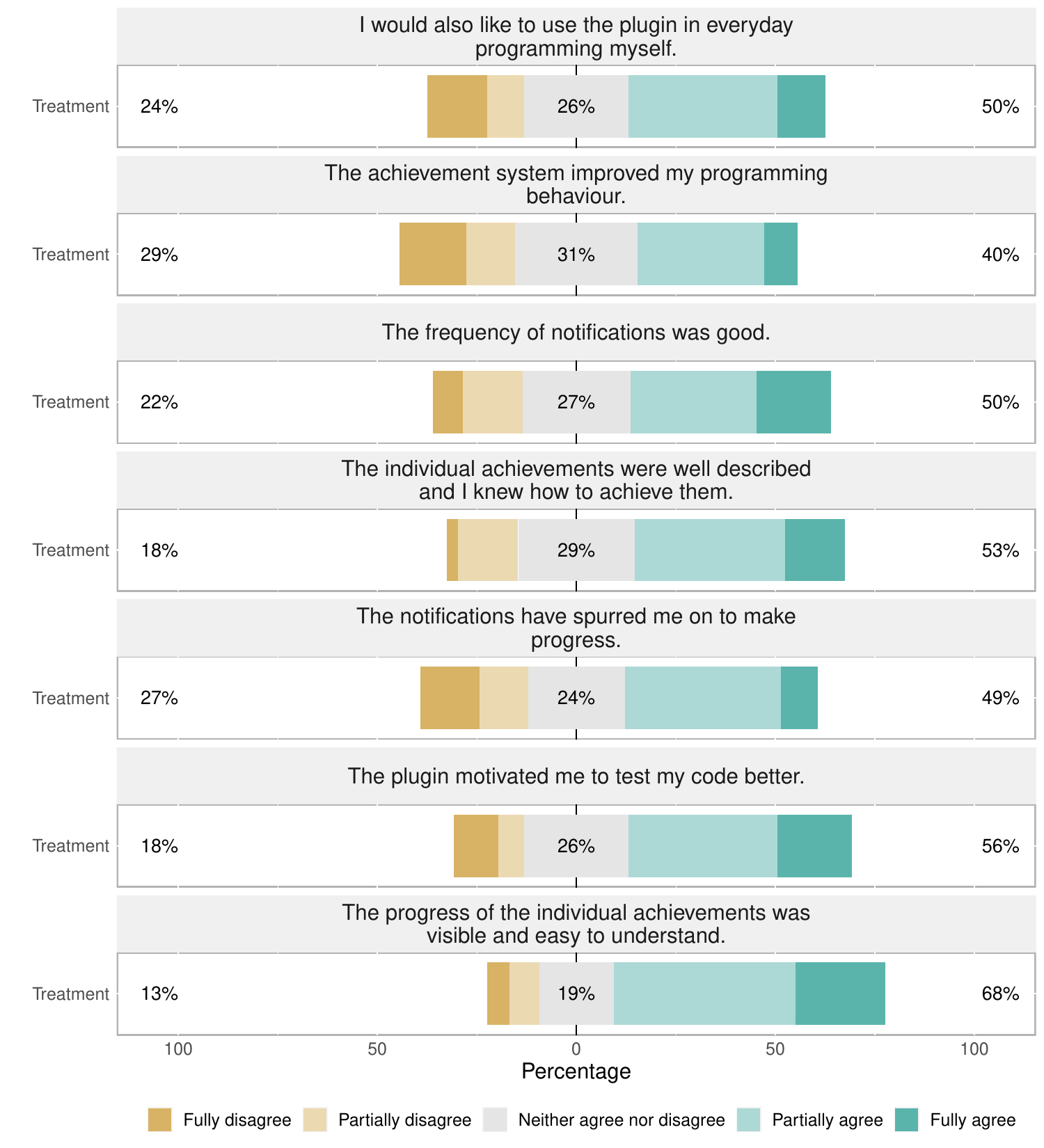}
        \caption{Answers on the \toolname plugin---ranging from negative on the left to positive on the right}
        \label{fig:exitsurveyplugin}
    \end{subfigure}

\caption{Survey responses of the participants as Likert plots}
\end{figure}

 \subsection{RQ5: Does \toolname influence the developer experience?}

According to the exit survey (\cref{fig:exitsurveyall}), the chosen target classes were appropriate, as participants from both groups reported having sufficient time to implement and test the class, and found it easy to understand and implement. Notably, fewer participants in the \treatment group felt they had enough time to complete testing (46\%) compared to the \control group (59\%), which is a significant difference ($p=0.013$). This uncertainty among the \treatment group is also reflected in their assessment of their test suite quality: only 27\% of the \treatment group believed they produced a good test suite, versus 38\% in the \control group, with this difference being nearly statistically significant ($p=0.058$). A possible explanation is that \toolname users were more aware of gaps in their testing due to the achievement prompts and coverage reports, which the \control group engaged with less often. This gave \toolname users a broader understanding of their test's quality and areas for improvement, whereas developers without \toolname are more likely to overestimate their testing efforts.

Among the participants of the \treatment group who used the achievements (\cref{fig:exitsurveyplugin}), 40\% felt that \toolname improved their programming habits, and about 50\% were motivated by notifications, including both encouragement and progress updates. Most participants understood the achievements and how to attain them, and around half expressed interest in using \toolname in their regular work. Over 50\% stated that \toolname motivated them to improve their testing practices, indicating that the plugin was effective in its goal.

\summary{RQ 5}{\toolname effectively motivated the \treatment group to improve testing practices and programming habits, with participants gaining greater awareness of test quality and areas for improvement, despite feeling slightly more time-constrained than the \control group.}

\comparison{Our findings confirm that \toolname encourages participants to conduct more testing; however, unlike in the original study, the participants of the \treatment group were less confident in their testing results compared to the \control group.}

\subsection{RQ6: Do users of \toolname write high-quality tests?} \label{sec:rq6}

\begin{lstlisting}[float, language=TypeScript, caption={Failing example test to check if the date is not in the past}, captionpos=b, label=lst:test1, xleftmargin=.5cm, xrightmargin=.1cm]
test("Datetype - False", () => {
    const result = isPast(new Date(2024, 6, 13));
    expect(result).toBe(false);
});
\end{lstlisting}

\begin{lstlisting}[float, language=TypeScript, caption={Failing example test to check if the date is in the future}, captionpos=b, label=lst:test2, xleftmargin=.5cm, xrightmargin=.1cm]
test("Datetype - True", () => {
    const result = isFuture(new Date(2024, 7, 13));
    expect(result).toBe(true);
  });
\end{lstlisting}

\begin{lstlisting}[float, language=TypeScript, caption={Failing example test because of wrong asumption of date format}, captionpos=b, label=lst:test3, xleftmargin=.5cm, xrightmargin=.1cm]
test("Add 10 days to 11/9 as string", ()=> {
    const test = new Date("9/11/2024");
    const result= addDays(test,10);
    expect(result.toLocaleDateString()).toBe('21/9/2024');
  })
\end{lstlisting}

\begin{lstlisting}[float, language=TypeScript, caption={Test smells for using the console rather than an assert statement}, captionpos=b, label=lst:smell, xleftmargin=.5cm, xrightmargin=.1cm]
console.log(isEqual(new Date(2024, 1, 22), new Date(2023, 1, 22)));
\end{lstlisting}

After analyzing the participants' test suites, we found that the test smell ratio was consistently higher in the \treatment group, with an average of 0.13 compared to the \control group's 0.03, a statistically significant difference ($p=0.013$, see \cref{fig:smells}). One commonly observed test smell is shown in \cref{lst:smell}, where a participant logged output to the console rather than using a Jest assertion. The significantly higher smell ratio in the \treatment group suggests that participants may have been more focused on writing a high number of tests---perhaps to gain achievements---than on test quality. An obvious solution to counter this problem would be to enhance \toolname with a new category of test achievements related to test smells. \RB{However, upon examining the specific test smells, we identified a total of 108 instances in the participants' test code. Of these, 101 were console outputs, while the remaining seven involved conditional statements within tests. Further analysis revealed that many console statements appeared alongside assertions, suggesting they were used for testing rather than debugging but were not removed afterward. This indicates that the overall test quality may not be as poor as suggested by Smelly Test.}

\RB{In RQ2 we noticed a substantial number of failing tests, when calculating mutation scores and therefore also investigated test robustness as an indicator of test quality. 
While the \treatment group had a significantly higher mean number of failed tests (1.6) compared to the \control group’s 0.73 ($p=0.037$, cf. \cref{fig:failed}), to some extent this is influenced by the increased number of tests written by participants of the \treatment group. The mean ratio of failing tests over tests written in total per participant is not significantly higher (\control: 0.062, \treatment: 0.114, $p=0.095$).} 

One possible explanation for failing tests might be brittle tests due to the use of a date-time library as target, which presents specific robustness challenges. Date and time handling requires attention to details like future test execution, various time zones, and date formats, yet these factors were often overlooked. For example, one participant’s test (\cref{lst:test1}) assumed a date would not be in the past—an assumption valid only during the experiment, leading to failures soon after. Another test in \cref{lst:test2} assumed a date should always be in the future, which was also invalidated shortly after the experiment ended. Additionally, in \cref{lst:test3}, a participant’s test expected output in \texttt{DD/M/YYYY} format instead of the system’s \texttt{DD/MM/YYYY}. \RB{While such tests may have passed at the time the experiment was conducted, they failed during our later analysis. These cases suggest that test robustness with respect to date and time handling was not something the participants were particularly aware of.}

\RB{To determine the effect of date-related problems, we re-ran the tests for each participant, this time adjusting the system time to match the commit time of the respective participant. Out of 214 initially failing tests, 61 passed with this adjustment, leaving 153 tests still failing due to other reasons. This reduction in failures is statistically significant ($p=0.014$), and the ratios of failing tests are reduced to 0.042 (\control) and 0.084 (\treatment), still with no significant difference between the two groups ($p=0.36$). Since brittle tests like in this scenario are not reported by common test smell detection tools, it may be worthwhile to investigate additional achievements for \toolname that explicitly reward robustness in tests.}

\RB{To further analyze the remaining failures, we manually reviewed the still-failing tests and identified three primary causes: (1) incorrect expected values in assertions, (2) incorrect implementations, and (3) incorrect date output formats. Since most participants with failing tests had only one or two failures, it is likely that they simply ran out of time at the end of the session and were unable to fully fix their tests or implementations. However, we suspect one other possible reason for the slightly higher failure rate in the \treatment group could be \toolname, as it rewards the number of tests and test executions rather than the number of \emph{passing} tests. Future work could therefore enhance \toolname and its achievements to address this limitation in the future.}

\summary{RQ 6}{More test smells and lack of robustness indicate that the \treatment group may focus on quantity over quality in their testing approach. Additional achievements in \toolname could be introduced to avoid these effects.}

\subsection{RQ7: What are the most important achievements?}

\begin{table}[]
    \caption{Mean progress and p-values of all achievements available in \toolname. Detailed information about the achievements can be found in the original paper~\cite{original_experiment}.}
    \label{tab:achievements}
    \centering
    \scriptsize
    \begin{tabular}{lrrr @{\hspace{1.2cm}} lrrr}
        \toprule
        \multirow{2}{*}{Achievement} & \multicolumn{2}{c}{Mean} & \multirow{2}{*}{p-value} &
        \multirow{2}{*}{Achievement} & \multicolumn{2}{c}{Mean} & \multirow{2}{*}{p-value} \\
                                     & Control & Treatment &       &
                                     & Control & Treatment &       \\ \midrule
        Test Executor                & 52.21   & 173.16     & \textbf{0.031}           &
        Check your classes           & 0.02    & 10.42      & \textbf{\textless 0.001} \\
        The Tester                   & 29.15   & 49.34      & 0.066                    &
        Check your branches          & 0.66    & 11.52      & \textbf{0.022}           \\
        The Tester - Advanced        & 0       & 0          & -                        &
        Class Reviewer - Lines       & 0.01    & 1.31       & \textbf{\textless 0.001} \\
        Assert and Tested            & 19.67   & 32.2       & 0.69                     &
        Class Reviewer - Methods     & 0       & 0          & -                        \\
        Bug Finder                   & 1.03    & 1.64       & 0.79                     &
        Class Reviewer - Branches    & 0       & 0          & -                        \\
        Safety First                 & 16.67   & 17.19      & 0.63                     &
        The Debugger                 & 1.25    & 2.09       & 0.087                    \\
        Gotta Catch 'Em All          & 0.13    & 2.34       & \textbf{\textless 0.001} &
        Take some breaks             & 2.31    & 8.29       & 0.43                     \\
        Line-by-Line                 & 2.03    & 52.12      & \textbf{\textless 0.001} &
        Make Your Choice             & 0       & 0          & -                        \\
        Check your methods           & 0.47    & 8          & \textbf{0.022}           &
        Break the Line               & 2.31    & 8.29       & 0.43                     \\
        Double check                 & 17.87   & 19.12      & 0.73                     &
                                     &         &            &                           \\ \bottomrule
    \end{tabular}
\end{table}

\Cref{tab:achievements} displays the achievements of \toolname developed for TypeScript, excluding those that are specific to Java. The table shows the average achievement actions for both the \treatment and \control groups, along with their p-values. An achievement action refers to any activity that contributes to progress toward the next achievement level. For instance, each time a participant uses the debugger, it counts towards the \textit{Debugger} achievement, and each combination of a failing test followed by a modification to the source code contributes to the \textit{Bug Finder} achievement.

The mean achievement levels were consistently higher in the \treatment group across various categories. A large and significant difference is observable for test executions, and achievements related to test coverage were also notably more advanced in the \treatment group, aligning with the observation that the \control group did not run tests with coverage reports enabled. The three achievements with the highest number of actions in the \treatment group were \textit{Test Executor} (for executing tests), \textit{The Tester} (for running test suites), and \textit{Line-by-Line} (for covering lines with tests). This indicates that simpler actions during testing were also the most influential for achievement progress. These achievements likely encouraged participants to write tests, execute them, and increase line coverage, as demonstrated in \cref{sec:rq1} and \cref{sec:rq2}.

Although there was significant progress on \textit{Class Reviewer - Lines}, overall advancement in the \textit{Class Reviewer} achievements was limited.
%
%
While this might be another example of effects that would require longer periods of usage to manifest, an alternative conjecture is that this is influenced by the parameterization of the achievements. For example, the level thresholds set in our experiment may have prevented developers from progressing, thus limiting the potential of these achievements. Although we used the original \toolname thresholds, future work should re-examine these parameters to provide stronger empirical evidence for setting optimal values.

Indeed some achievements are designed for long-term engagement, which may not be observable within the timeframe of our experiment. For instance, achievements like \textit{The Tester - Advanced} require running at least ten test suites with a minimum of 100 tests---an expectation that was not feasible within our experiment’s duration. Thus, these types of achievements may be more appropriate for long-term application rather than short-term studies.
This highlights the need for further replicability studies that examine the effects of gamification over extended periods. 

The \textit{Make Your Choice} achievement, which requires setting conditional breakpoints, was not utilized by any participants in the \treatment group. While it is understandable that the \control group did not engage with this feature due to its obscurity, the lack of attempts by the \treatment group is unexpected.
%
More generally we observe low pursuit of debugging achievements, reflected in their low mean scores in \cref{tab:achievements}. The \textit{Bug Finder} achievement, which rewards participants for modifying source code following a failing test, highlights the difficulty developers face with debugging tasks. While the mean achievement level was higher in the \treatment group, this difference was not statistically significant compared to the \control group. This raises questions about the effectiveness of certain debugging achievements, especially those tied to specific types of breakpoints. Future iterations of the experiment might consider introducing new achievements that are easier to attain to better incentivize developers to engage in debugging behaviors. 


Achievements such as \textit{Assert and Tested} and \textit{Double Check} were completed at similar rates by both groups, suggesting they may not have a strong impact in their current form. This calls for a re-evaluation of their design or reward structure to more effectively encourage developers to incorporate assertions in their tests in future implementations.

Finding the right balance between easily obtainable and more challenging achievements warrants further investigation. Given the lower completion rates for achievements like \textit{The Tester - Advanced} and those related to debugging, introducing additional, intermediate achievements could help guide developers gradually toward more complex behaviors. Providing other forms of support, such as hints on how to achieve complex goals, could also be valuable for developers.

\summary{RQ7}{While certain achievements in \toolname for TypeScript enhance user engagement and promote best practices in testing, the effectiveness of long-term achievements and debugging incentives varies, highlighting the need for further studies to explore their impact over extended periods and refine the achievement system.}

	\section{Discussion}

\RC{The original study found a significant improvement in branch coverage when users were exposed to gamification.  In our experiment, although the treatment group produced slightly higher code coverage values, the differences between the \control and \treatment groups are not statistically significant. An explanation for this result might be the fact that inherently, the Date-Fns project does not have many branches by nature. Most of the functions indeed can be implemented with few lines of code and with a very limited number of conditional statements, therefore the variability of the measured branch coverage is limited. Future replicability studies may want to consider fewer, but more complex functions as experiment tasks, to allow for larger variations.}

\RB{Besides coverage differences, in the original study test effectiveness was significantly improved for users creating the test suite with gamification, as quantified by the mutation score. In contrast, in this study we did not find a significant difference in this metric. Upon further examination of the mutants produced by the Stryker mutation tool, we noticed that for most of the functions to be implemented, the generated mutants were largely redundant (e.g., several mutants modifying the modulus operator with a multiplication). This redundancy can also be attributed to the inherent simplicity of certain functions, which, lacking complex logic, leads to the generation of redundant mutants that ultimately result in similar mutation scores.}

\RA{Contrary to the original study, our participants in the gamified environment reported more time pressure and slightly less confidence in their test results. 
This might be influenced by the continuous feedback loop of achievement notifications and progress indicators, 
which may have raised awareness of test quality in general. In particular, if thresholds for achievement levels are set too high, then it would be conceivable that users would feel pressured to invest too much time into testing until the next level is reached. Consequently, it would be useful to revisit the thresholds currently set in \toolname, which have not yet been systematically explored and optimized.}

\RA{The perceived time pressure may also have led some participants to prioritize quantity over quality, as seen in higher test smell ratios and increased failure rates in the \treatment group. These findings suggest that, while gamification enhances testing engagement, there is potential to further refine achievement structures to foster both quality and quantity in test creation. An important oversight in the existing achievement system of \toolname revealed by our study is that failing tests count just like passing ones when \toolname counts the events that lead to awarding an achievement. However, adding failing tests should arguably only be rewarded if that test failure reveals an actual bug which is successively fixed. It is also conceivable to introduce further achievements related to test quality (e.g., awarding testers that remove test smells.)}

\RB{Since the \treatment group began testing earlier with the help of \toolname, they were able to identify and fix more bugs early on, whereas the \control group initially neglected testing. Despite similar coverage and mutation scores, the higher number of tests in the \treatment group suggests they spent more time detecting bugs early. This indicates that \toolname played a crucial role in improving implementation correctness rather than the difference being purely coincidental.}

\section{Conclusions}

\label{sec:conclusions} 

Our replication study confirms that gamification, through the \toolname plugin, significantly influences software testing behaviors in an IDE setting, supporting findings from the previous study and extending them with new insights. Gamification in the \treatment group led to more frequent test creation, increased use of coverage tools, and engagement with debugging features, demonstrating \toolname's ability to encourage more rigorous testing practices. Notably, the \treatment group also achieved a greater number of tests, indicating enhanced attention to finer details in test cases. These behaviors correlate with improved test suite quality metrics, higher achievement levels, and earlier functionality in the codebase.

Future research could explore adjustments to gamification frameworks within IDEs, focusing on balancing achievements that emphasize quality with those targeting quantity. Specifically, introducing achievements that reward robust test creation—such as those that prevent common test smells or require robust handling of time-sensitive data—could help improve the depth and reliability of testing practices. \RC{IntelliGame’s incentive mechanism can be extended to balance test quality and quantity more effectively. Enhancing enforcement mechanisms for syntactic correctness, such as integrating a type-checking module or providing real-time feedback, could improve test input validity. Additionally, incorporating branch coverage visualization may help users construct more effective test cases.} \RCremoved{Additionally} \RC{Furthermore}, longer-term studies with larger sample sizes could help determine whether gamification's influence on testing behavior extends beyond the immediate experimental context and persists in professional development settings. Lastly, incorporating adaptive gamification, where the plugin tailors feedback based on individual tester progress, could further optimize user engagement and development outcomes.
	
	\section{Data Availability}
	The artifacts are available at \url{https://doi.org/10.6084/m9.figshare.27443505}

	\begin{acks}
		
		This study was carried out within the “EndGame - Improving End-to-End Testing of Web and Mobile Apps through Gamification” project (2022PCCMLF) – funded by European Union – Next Generation EU within the PRIN 2022 program (D.D.104 - 02/02/2022 Ministero dell’Università e della Ricerca). This manuscript reflects only the authors’ views and opinions and the Ministry cannot be considered responsible for them.
		
	\end{acks}
	
	\newpage
	
	\bibliographystyle{ACM-Reference-Format}
	\bibliography{bib}
	
\end{document}